\documentclass[fleqn,usenatbib]{mnras}
\usepackage{newtxtext,newtxmath}
\usepackage[T1]{fontenc}
\usepackage{ae,aecompl}
\usepackage{graphicx}	
\usepackage{amsmath}
\usepackage{times}
\usepackage{verbatim}
\usepackage{subfigure}
\usepackage{float}
\usepackage{color}

\usepackage{amssymb}

\title[High CGM angular momentum keeps galaxy quenched]{From large-scale environment to CGM angular momentum to star forming activities -- II: quenched galaxies}

\author[S. Lu et al.]
{Shengdong Lu$^{1}$\thanks{E-mail: \url{lushengdong@tsinghua.edu.cn}},
Dandan Xu$^{1}$\thanks{E-mail: \url{dandanxu@tsinghua.edu.cn}},
Sen Wang$^{1}$,
Zheng Cai$^{1}$,
Chuan He$^{2,3,4}$,
C. Kevin Xu$^{2,4}$,
Xiaoyang Xia$^{5}$,
\and 
Shude Mao$^{1,2}$,
Volker Springel$^{6}$,
Lars Hernquist$^{7}$
\\
\\
$^{1}$Department of Astronomy, Tsinghua University, Beijing 100084, China\\
$^{2}$National Astronomical Observatories, Chinese Academy of Sciences, Beijing 100101, China\\
$^{3}$School of Astronomy and Space Sciences, University of Chinese Academy of Sciences, Beijing 100049, China \\
$^{4}$Chinese Academy of Sciences South America Center for Astronomy, Beijing 100101, China \\
$^{5}$Tianjin Astrophysics Center, Tianjin Normal University, Tianjin 300387, China\\
$^{6}$Max-Planck-Institut f\"ur Astrophysik, Karl-Schwarzschild-Str. 1, D-85748, Garching, Germany\\
$^{7}$Harvard-Smithsonian Center for Astrophysics, 60 Garden Street, Cambridge, MA 02138, USA
}

\date{Accepted ***. Received ***; in original form ***}

\pubyear{2021}

\begin{document}
\label{firstpage}
\pagerange{\pageref{firstpage}--\pageref{lastpage}}
\maketitle
\begin{abstract}
The gas needed to sustain star formation in galaxies is supplied by the circumgalactic medium (CGM), which in turn is affected by accretion from large scales. In a series of two papers, we examine the interplay between a galaxy's ambient CGM and central star formation within the context of the large-scale environment. We use the IllustrisTNG-100 simulation to show that the influence exerted by the large-scale galaxy environment on the CGM gas angular momentum results in either enhanced (Paper I) or suppressed (Paper II, this paper) star formation inside a galaxy. We find that for present-day quenched galaxies, both the large-scale environments and the ambient CGM have always had higher angular momenta throughout their evolutionary history since at least $z=2$, in comparison to those around present-day star-forming disk galaxies, resulting in less efficient gas inflow into the central star-forming gas reservoirs. A sufficiently high CGM angular momentum, as inherited from the larger-scale environment, is thus an important factor in keeping a galaxy quenched, once it is quenched. The process above naturally renders two key observational signatures: (1) a coherent rotation pattern existing across multiple distances from the large-scale galaxy environment, to the circumgalactic gas, to the central stellar disk; and (2) an anti-correlation between galaxy star-formation rates and orbital angular momenta of interacting galaxy pairs or groups.
\end{abstract}


\section{Introduction}
\label{sec:introduction}
The presence (or absence) of active star formation is one of the most distinctive characteristics of a galaxy. Many other descriptive properties, such as galaxy color, stellar age etc., are simply consequences of it. How once star-forming galaxies transition into a quenched state is thus a key question in understanding galaxy formation. For example, the (initial) formation of massive quenched elliptical galaxies is thought to be closely related to major mergers, during which their cold star-forming gas reservoirs are largely and quickly consumed by triggered starbursts, supermassive black hole accretion and associated feedback effects, resulting in quenched stellar bulges at the centres of galaxies (e.g., \citealt{Barnes_et_al.(1992),Barnes_et_al.(1991),Barnes_et_al.(1996),Mihos_et_al.(1996), Barnes_et_al.(2002),Naab_et_al.(2003),Bournaud_et_al.(2005),Bournaud_et_al.(2007),Johansson_et_al.(2009a),Tacchella_et_al.(2016),Rodriguez-Gomez_et_al.(2017)}). Additional quenching mechanisms (for a central galaxy in a dark matter halo) include virial shocks (also known as halo quenching; \citealt{Birnboim_et_al.(2003),Dekel_et_al.(2006),Birnboim_et_al.(2007), Keres_et_al.(2009)}), AGN feedback (e.g., \citealt{Granato_et_al.(2004),Hopkins_et_al.(2005),Di_Matteo_et_al.(2005),Di_Matteo_et_al.(2008),Croton_et_al.(2006),Cattaneo_et_al.(2009),Johansson_et_al.(2009b)}), and morphological quenching~\citep{Martig_et_al.(2009)}. 

In this series of two papers, we consider star formation and quenching activities within the context of a galaxy's ambient and environmental angular momenta. Specifically we use the IllustrisTNG-100 simulation~\citep{Marinacci_et_al.(2018),Naiman_et_al.(2018),Nelson_et_al.(2018),Pillepich_et_al.(2018b),Springel_et_al.(2018)} to investigate modulations of star formation induced by the circumgalactic medium (CGM), which inherits angular momentum from the galaxy's large-scale environment. In Paper I of this series \citep{Wang_et_al.(2021)}, we focus on present-day star-forming disk galaxies and investigate their typically episodic star-formation activities, intermittent with quenched phases. In this paper, we study these effects for present-day {\it quenched} galaxies. We find that the two types of angular momenta at two different distance scales, i.e., the larger-scale environmental angular momentum (from the orbital motion of neighbouring galaxies) and the CGM spin are closely correlated: the latter experiences a modulation from the former through galaxy interactions. Higher CGM angular momentum is observed to correlate with lower star-formation rates as well as lower AGN accretion rates. These relations are independent of galaxy mass and can be well explained by the fact that sufficiently high CGM angular momentum would result in less efficient gas inflow into the central star-forming gas reservoirs. In the case of present-day quenched elliptical galaxies, this would {\it keep} them quenched once they become quenched in the first place (due to other mechanisms as listed above).

The connections between galaxies' large-scale environments, their CGM angular momenta and star forming/quenching activities have significant impacts on the different fates of galaxies. In the second half of this paper, we specifically compare three types of galaxies; i.e., present-day star-forming disk galaxies, and fast- and slowly-rotating early-type galaxies. The latter two have been observed and extensively studied by the SAURON project (e.g. \citealt{Cappellari_et_al.(2007),Emsellem_et_al.(2007)}) and the ATLAS$^{\rm 3D}$ survey (e.g. \citealt{Emsellem_et_al.(2011), Cappellari_et_al.(2013)}). In particular, we take the galaxy samples from \citet{Lu_et_al.(2021b)}, where the simulation counterparts of the three above-mentioned types of galaxies (with stellar masses of $10.3\,\leqslant\,\log\,M_{\ast}/\mathrm{M_{\odot}}\,\leqslant\,11.2$) were identified in the IllustrisTNG-100 simulation. We find that both the orbital angular momenta of neighbouring galaxies and the fractions of tangential motion of the CGM gas are the highest among slowly-rotating ellipticals, moderately high among fast-rotating early-type galaxies, and the lowest among star-forming disk galaxies. This can explain the quenched states of fast- and slowly-rotating early-type galaxies, with respect to their star-forming disk counterparts. For the latter, the CGM angular momenta are just large enough to maintain their rotational stellar and gaseous disks, but not large enough to prevent efficient gas inflow to fuel star formation and feed the supermassive black holes at the centres of galaxies. In contrast, quenched early-type galaxies, with {\it higher} CGM angular momenta, would thus {\it not} receive sufficient (inner) gas replenishment. They can therefore easily remain quenched once their central star formation activities cease. For fast rotators, in particular, once they have consumed the majority of their inner gas reservoirs, star formation may only gently carry on in the galaxy outskirts, where the cold gas spirals about. This is strongly supported by the frequent appearance of gaseous ring features at larger radii of these galaxies (also see \citealt{Lu_et_al.(2021b)}).

It is worth noting that in comparison to star-forming disk galaxies, the environmental angular momenta and the CGM gas spins around quenched early-type galaxies have always been higher since as early as $z\sim 2$. This indicates that their different fates might already have been fixed by initial conditions at their birth sites. The initial conditions determine the large-scale torque fields, hence they map out at some level the coarse-grained ``paths'' of later-stage galaxy interactions. The CGM gas then inherits the associated orbital angular momentum of interacting galaxies in the environments. An efficient (CGM) gas inflow from larger distances into the galaxy would then be related to the CGM angular momentum. Such a modulation (together with other quenching mechanisms) can directly affect the star-forming or quenching status of a galaxy.

In line with this scenario, the TNG simulation predicts a coherent rotation pattern existing among: (1) the large-scale environment out to $\sim 100$ kpc, as directly measured by neighbouring galaxies' line-of-sight velocities; (2) the cold (with ${\rm T}<2\times 10^4\,{\rm K}$) CGM gas, also extending to several tens of kpc; and (3) the central stellar disk of a few kpc. Such coherent kinematics can be obtained through stacking the line-of-sight velocity fields of galaxies that exhibit cold stellar dynamics. Specifically, spectroscopic measurements of neighbouring galaxies are required to be able to resolve relative line-of-sight orbital motion below 100 km/s. The CGM gas kinematics can be obtained from absorption line observations towards quasar sight lines which go through the CGM of foreground galaxies. Such observations at multiple distance scales could provide strong evidence about the angular momentum inheritance from the large-scale environment at several tens of kpc scales, through the CGM, down to central stellar disks at kpc scales. We also propose to seek for anti-correlations between galaxy star-formation rates and orbital angular momenta of nearby interacting galaxies. 

This paper is organized as follows. In Section~\ref{sec:data}, we introduce the IllustrisTNG-100 galaxy samples that we adopt for this study. The connection between the CGM gas spin and the central star-forming activities, and that between the environmental angular momentum and the CGM gas spin are examined in Section~\ref{sec:CGMvsSF} and \ref{sec:FromLSOAMtoCGM}, respectively. We discuss the implications for different fates of galaxies in Section~\ref{sec:FATES}. Key observational predictions are presented in Section~\ref{sec:observation}. Finally, conclusions and some further discussion are given in Section~\ref{sec:conclusion}.
In this work, we adopt the Planck cosmology \citep{Planck_Collaboration(2016)}, which was also used in the IllustrisTNG-100 simulation. In particular, a flat universe geometry is assumed, with a total matter density of $\Omega_{\rm m} = 0.3089$, a cosmological constant of $\Omega_{\Lambda} = 0.6911$, a baryonic density of $\Omega_{\rm b} = 0.0486$, and a Hubble constant $h = H_0/(100\,{\rm km s}^{-1} {\rm Mpc^{-1}}) = 0.6774$.

\section{Methodology}
\label{sec:data}

\subsection{Galaxy samples from the IllustrisTNG Simulation}
\label{sec:tng}
\textit{The Next Generation Illustris Simulations}\footnote{\url{http://www.tng-project.org/data/}} (IllustrisTNG, TNG
hereafter; \citealt{Marinacci_et_al.(2018),Naiman_et_al.(2018),Springel_et_al.(2018),Nelson_et_al.(2018),Nelson_et_al.(2019b),Pillepich_et_al.(2018a),Pillepich_et_al.(2018b),Pillepich_et_al.(2019), Weinberger_et_al.(2017),Weinberger_et_al.(2018)}) are a suite of state-of-the-art magneto-hydrodynamic cosmological galaxy formation simulations carried out in large cosmological volumes with the moving-mesh code \textsc{arepo} \citep{Springel(2010)}. In addition to the previous galaxy formation model (see \citealt{Genel_et_al.(2014),Vogelsberger_et_al.(2013),Vogelsberger_et_al.(2014b),Vogelsberger_et_al.(2014a),Nelson_et_al.(2015)}), the simulation has also adopted ideal magneto-hydrodynamics, included a new model for AGN feedback \citep{Weinberger_et_al.(2017),Weinberger_et_al.(2018)}, and implemented various modifications to galactic wind feedback, stellar evolution and chemical enrichment \citep{Pillepich_et_al.(2018a)}. 

We use the full-physics version with a cubic box of $110.7\,\mathrm{Mpc}$ side length (TNG-100), a gravitational softening length of $0.5h^{-1}\,\mathrm{kpc}$ and mass resolution of $1.4\times 10^6\,{\rm M_{\odot}}$ and $7.5\times10^6\,{\rm M_{\odot}}$ for the baryonic and dark matter, respectively. The {\sc subfind} algorithm \citep{Springel_et_al.(2001),Dolag_et_al.(2009)} has been used to identify galaxies in their host dark matter halos. 

In this work, three galaxy samples within a stellar mass range of $10.3\,\leqslant\,\log\,M_{\ast}/\mathrm{M_{\odot}}\,\leqslant\,11.2$ at $z=0$ are used. These galaxy samples were originally selected from the TNG-100 simulation and used in \citet{Lu_et_al.(2021b)} in order to study the evolution of bulge-dominated quenched but dynamically cold galaxies (referred to as CQs denoting ``cold quenched''), in comparison to quenched and dynamically-hot elliptical galaxies (referred to as NEs denoting ``normal elliptical'') and star-forming disk galaxies (referred to as NDs denoting ``normal disk''). In terms of star-forming activity, morphology and kinematics, the cold quenched and hot quenched early-type galaxies were shown to match the observed fast- and slowly-rotating early-type galaxies, respectively (see Figure 10 in \citealt{Lu_et_al.(2021b)}). In this work, we follow the same name conventions and use the corresponding notations in figures.

\subsection{Galaxy and environmental properties}
\label{sec:def}
General properties of the simulated galaxies are taken from the TNG public catalogue \citep{Nelson_et_al.(2019a)}. In addition, we also carry out calculations for a number of other key properties related to the galaxy itself, its CGM and its environmental angular momentum. To do so, we first define the galaxy domain as the region inside twice the half-stellar-mass radius $R_{\rm hsm}$ from the galaxy centre; and the region outside $2R_{\rm hsm}$ as the CGM domain \citep{DeFelippis_et_al.(2020)}. The relevant properties inside the galaxy domain include a galaxy's specific star formation rate (sSFR), central gas fraction $f_{{\rm gas},\,<2R_{\rm hsm}}$, and specific gaseous and stellar disk spin vectors $\bf{j_{\rm gas}}$ and $\bf{j_{\ast}}$, respectively. $\bf{j_{\rm gas}}$ and $\bf{j_{\ast}}$ are calculated as:
\begin{equation}
\label{eq:j}
    {\bf j}_{\it x} = \frac{\sum_i\, m_i\, \bf{r}_i \times \bf{v}_i}{\sum_i\, m_i},
\end{equation}
where the subscript $x$ refers to stars or gas inside the galaxy domain; $m_i$, $\it{\bf{r_i}}$ and $\it{\bf{v_i}}$, respectively, are the mass, position and velocity vectors with respect to the galaxy centre, of the $i$-th stellar/gaseous element. The summation goes over all the stellar/gaseous particles/cells within $2R_{\rm hsm}$. We refer the reader to \citet{Lu_et_al.(2021b)} and \citet{Xu_et_al.(2019)} for more detailed descriptions. 

For the CGM domain, we also calculate the specific angular momentum vector $\bf{j_{\rm CGM}}$ of a galaxy's CGM gas using Eq.~\ref{eq:j}, with the summation extending over all the gaseous cells outside $2R_{\rm hsm}$ that are gravitationally bound to the galaxy's halo. We verify that the results are not subject to the choice of the radial bounds that we use to calculate the CGM gas properties. The results remain unchanged when we confine our calculations to within a fixed radial range of 30-300 kpc from the galaxy centre.

In order to eliminate the effects of mass dependence, we also define a dimensionless spin for the CGM around each galaxy as follows:  
\begin{equation}
\it{\bf{\lambda_{\rm CGM}}} = \sum_i\, \frac{m_i\, v^2_{{\rm t},\,i}}{GM(\leqslant r_i) / r_i}\bigg/{\sum_i\, m_i},
\label{eq:lcgm}
\end{equation}
where $m_i$, $r_i$ and $v_{{\rm t},\,i}$, respectively, are the mass, distance and tangential velocity of the $i$-th gas element with respect to the galaxy centre, $G$ is the gravitational constant, and $M(\leqslant r_i)$ is the total mass enclosed within distance $r_i$. To evaluate and compare among different galaxy types the spin of the CGM as a whole, the summation is over all gas elements within a fixed radial range of 30-300 kpc from the galaxy centre. We note that the dimensionless spin defined in this way is different from the conventional definition (e.g., \citealt{Stewart11CGMaccretion, Stewart17CWstreamHiAM, Teklu15Spin, Danovich15CWstreamToDisk}), which is essentially a globally-defined angular momentum ratio evaluated within a certain radius. The definition used here is a mass-weighted energy fraction of the tangential motion with respect to the total kinetic energy that is required to balance gravity at this radius (assuming a spherically-symmetric matter distribution). The concern here is the following: as demonstrated in Paper I (see Figures 7 and 8 therein), the CGM gas shows a bimodal kinematics between the cold and hot phases. The cold gas, in particular, can exhibit rather localized filamentary morphologies with certain in-spiral motions, which are not necessarily global features. We have therefore adopted the definition above. The closer this value is to 1, the higher the fraction of gas is that is moving in a tangential flow, in comparison to being in a hydrostatic state supported by pressure gradients.

We have also investigated the angular momentum distribution of individual CGM gas elements in a given galaxy, through an angular momentum ratio $\epsilon$, which is similar to ``circularities'' defined for individual stars in a galaxy (e.g., see \citealt{Xu_et_al.(2019)} and \citealt{Buck_et_al.(2020)} for detailed definition). This gas ``circularity'' $\epsilon$ is defined as follows:
\begin{equation}
\epsilon = \frac{j_{\parallel}}{\sqrt{GM(\leqslant r)r}},
\label{eq:epsilon}
\end{equation}
where $j_{\parallel}$ is the specific angular momentum of the gas cell along the direction of the total CGM spin, and $r$ is the distance of the gas element; the denominator is essentially the maximum angular momentum allowed for a steady Kepler orbit at distance $r$ (however gas is not moving along Kepler orbit). In Section \ref{sec:GalDiffInCGMspin}, we compare the $\epsilon$ distributions of the CGM gas among different types of galaxies.

%note here that we have checked that the mass fraction of gas being retrograde (with respect to the total CGM spin direction). So the dimensionless spin so defined can be used to quantify the tangential motion of the CGM gas without regard to their rotation directions. We refer the reader to \citet{Xu_et_al.(2019)} and \citet{Buck_et_al.(2020)} for the calculation of the degree of co-rotation for both stars and gas.

In our study, all galaxies in the samples are central galaxies of Friends-of-Friends (FoF, \citealt{Davis_et_al.(1985)}) groups. Their environmental angular momenta are specifically measured by the specific orbital angular momentum vector $\it{\bf{j_{\rm env}}}$, which is given by:
\begin{equation}
\label{eq:jenv}
\it{\bf{j_{\rm env}}} = \frac{\sum_i\, M_{{\rm tot},\,i}\, \it{\bf{r_i}} \times \it{\bf{v_i}}}{\sum_i\, M_{{\rm tot},\,i}},
\end{equation}
where $M_{{\rm tot},\,i}$, $\it{\bf{r_i}}$ and $\it{\bf{v_i}}$, respectively, are the total mass, relative distance and velocity vectors with respect to the investigated central galaxy, of the $i$-th neighbouring galaxy (subhalo) in the vicinity. The summation is over all neighbouring galaxies, for which we take all satellite galaxies in the same FoF groups and within a given distance range from $\rm 50\,kpc$ to $\rm 300\,kpc$ measured from the central galaxy position. We again confirm that adopting the fixed distance range above or taking a radial range from one to eight times the half-{\it total}-mass radius to evaluate the environmental angular momentum makes no qualitative difference to the findings in this work.

\section{From CGM angular momenta to star formation and quenching activities}
\label{sec:CGMvsSF}

\subsection{The mass and spin of the inner and outer gas}
\begin{figure*}
\includegraphics[width=1.6\columnwidth]{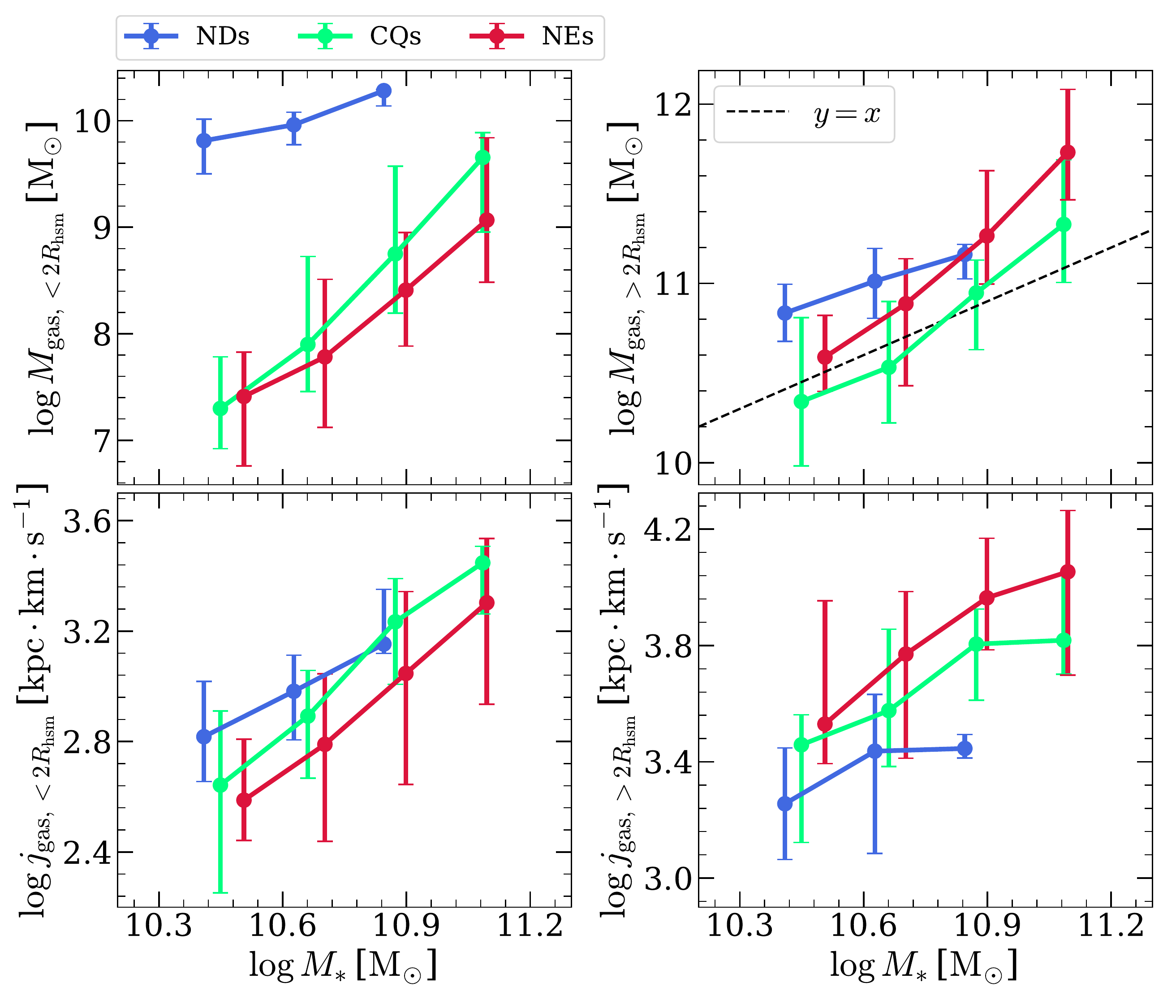}
\caption{Left: the distributions of mass (top) and specific spin (bottom) of inner ($r<2R_{\rm hsm}$) gas, as a function of galaxy stellar mass $M_{\ast}$, in cold quenched galaxies (green), normal elliptical galaxies (red), and normal disk galaxies (blue). Right: the same plots for the outer gas ($r>2R_{\rm hsm}$, also defined as the CGM domain). The error bars indicate the range from the 16th to the 84th percentiles ($\pm 1\sigma$). The black dashed line in the top right panel indicates $M_{\ast}=M_{\mathrm{gas},>2R_{\rm hsm}}$.}
\label{fig:gasmassspin}
\end{figure*}

To understand the link between the CGM angular momentum and the central star formation and quenching activities, we first present Fig.~\ref{fig:gasmassspin}, in which we show distributions of the mass $M_{\rm gas}$ and specific angular momentum $j_{\mathrm{gas}}$ of the inner gas (specifically denoted with subscript $<2R_{\rm hsm}$) and of the outer/CGM gas (denoted with subscript $>2R_{\rm hsm}$), as a function of galaxy stellar mass $M_{\ast}$. We note that by our definition, $j_{{\rm gas},\,>2R_{\rm hsm}}$ equals to $j_{\rm CGM}$. As can be seen, the masses and specific spins of both inner and outer/CGM gas increase with increasing $M_{\ast}$. As expected, star-forming disk galaxies host significantly more abundant gas inside the galaxy domains (left panels), and also exhibit higher gas spins, in comparison to their quenched galaxy counterparts within the same stellar mass range. 

However, the gas content and angular momentum in the CGM regime (panels on the right) behave differently from those within the galaxy domain. Specifically, differences in the CGM gas mass $M_{{\rm gas},\,>2R_{\rm hsm}}$ among the three types of galaxies become much smaller. In particular, the outer regions around quenched early-type galaxies are not as ``gas-poor'' as expected; $M_{{\rm gas},\,>2R_{\rm hsm}}$ is typically a few times higher than $M_{\ast}$ on average, suggesting that quenched early-type galaxies do not lack an {\it outer} gas supply, which may potentially fuel their central star formation (but which is not happening). More interestingly, the specific angular momenta of the CGM gas $j_{{\rm gas},\,>2R_{\rm hsm}}$ around star-forming disks are the lowest among the three types of galaxies, while quenched galaxies (within the same stellar mass range) exhibit higher CGM spins. In the next subsection, we show an important consequence of these differences.

\subsection{The connection between the CGM spin and star formation}
\label{sec:CGMvsSFR}

\begin{figure}
\includegraphics[width=1\columnwidth]{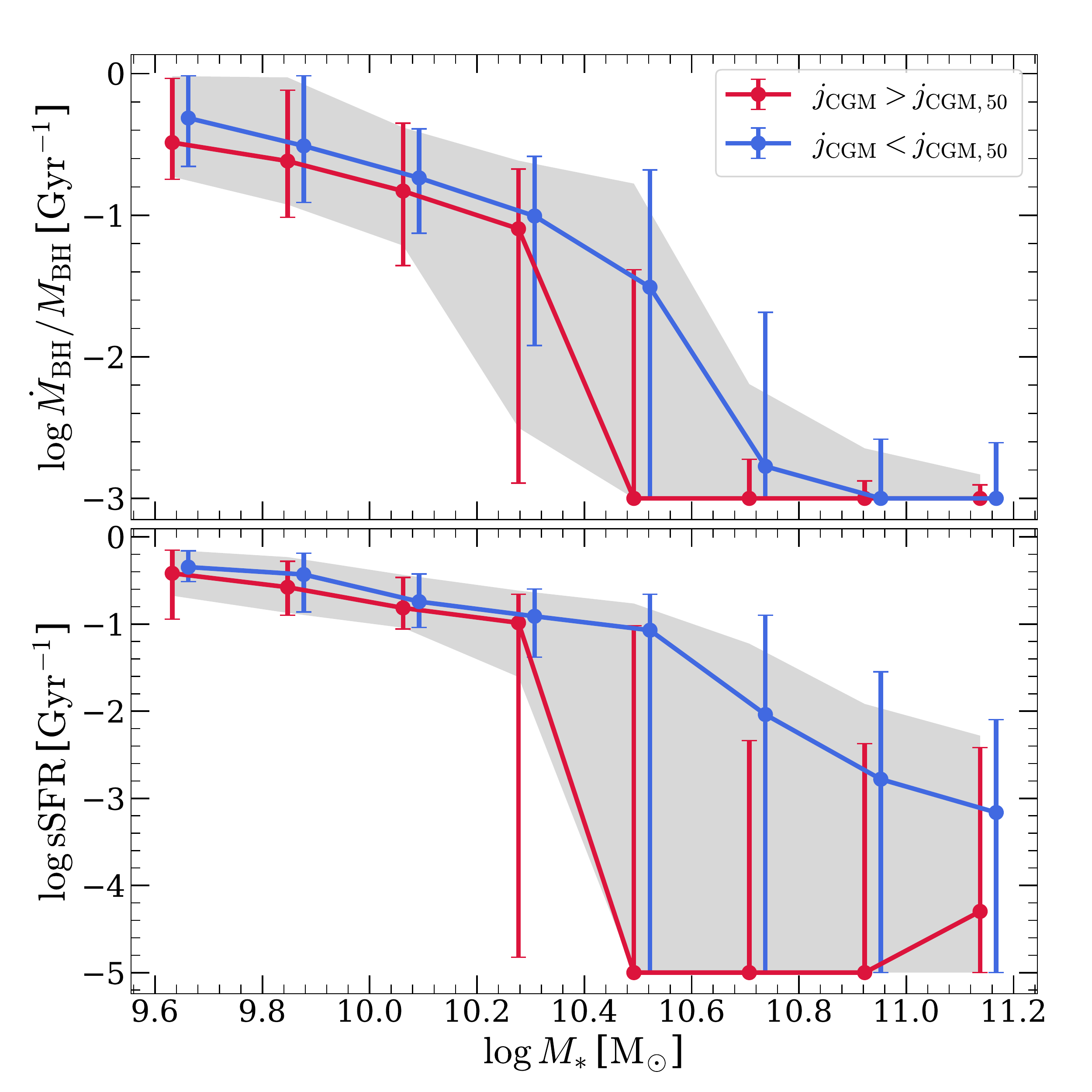}
\caption{The distributions of the black hole accretion rate $\dot{M}_{\rm BH}/M_{\rm BH}$ (top panel) and the specific star-formation rate $\mathrm{sSFR}$ within $2R_{\rm hsm}$ (bottom panel) as a function of galaxy stellar mass $M_{\ast}$ for all the three types of galaxies (i.e. NDs, CQs, and NEs) and their progenitors since $z = 1$. In each panel, galaxies are divided into 8 bins according to their stellar masses, with the grey shaded region indicating the $\pm 1\sigma$ region of all the galaxies. In each bin, galaxies are further divided into two subgroups according to their CGM angular momenta $j_{\rm CGM}$; i.e., galaxies with $j_{\rm CGM}>j_{\rm CGM,50}$ (red) and with $j_{\rm CGM}<j_{\rm CGM,50}$ (blue), where $j_{\rm CGM,50}$ is the median $j_{\rm CGM}$ in each stellar mass bin. The error bars indicate the range from 16th to 84th percentiles ($\pm 1\sigma$).}
\label{fig:CGMspinSFRAnticorrelation}
\end{figure}

\begin{figure}
\includegraphics[width=1\columnwidth]{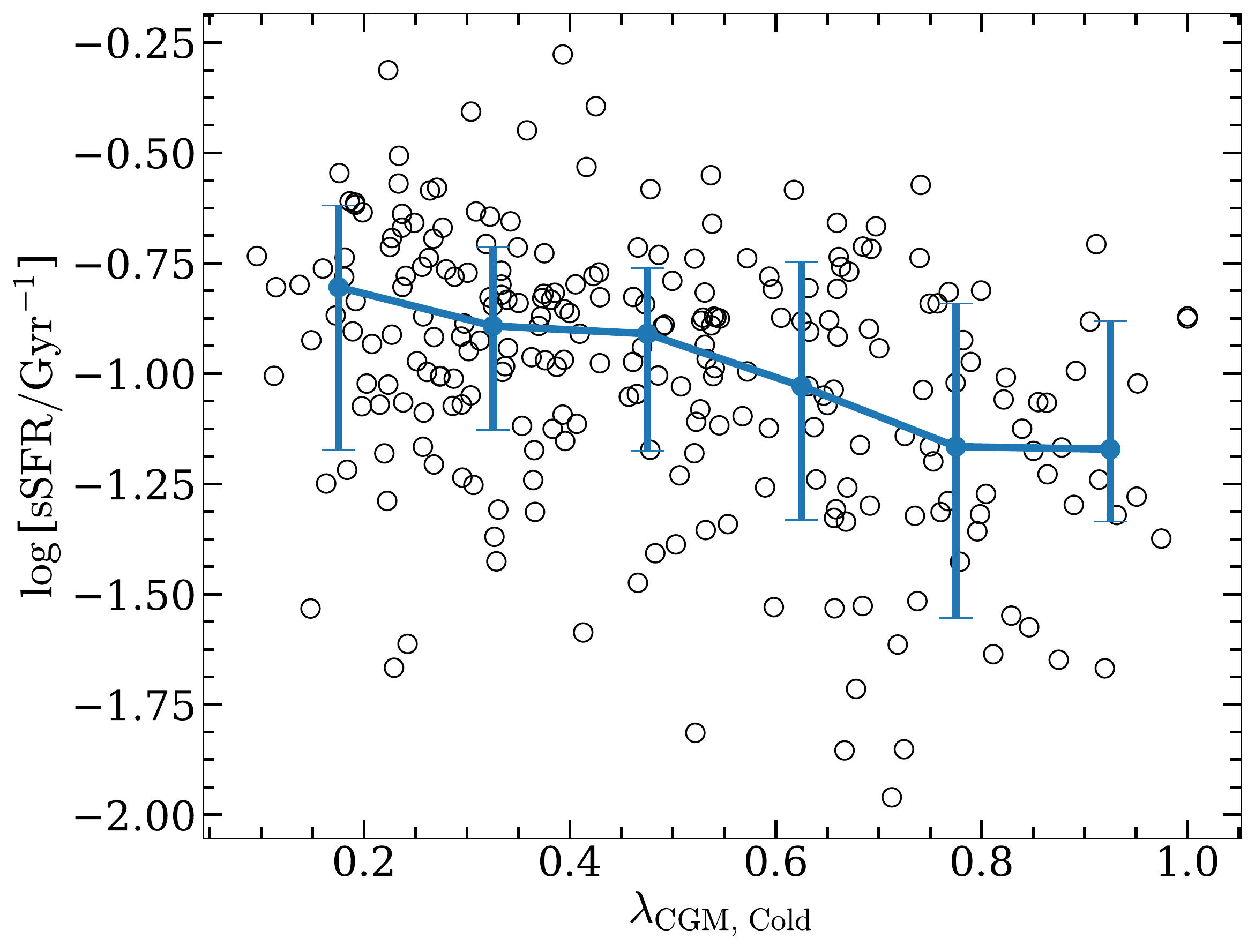}
\caption{The distribution of the logarithmic sSFR as a function of dimensionless spin $\lambda_{\rm CGM}$ of the cold CGM gas (i.e., with $10^4\,{\rm K} < {\rm T} < 2\times 10^4 \,{\rm K}$, and located within a radial range of $30-100$ kpc) for star-forming disk galaxies at $z\leqslant 0.1$. The solid line connects the median values in $\lambda_{\rm CGM}$ bins; the error bars indicate the ranges from 16th to 84th percentiles ($\pm 1\sigma$) of the distribution. }
\label{fig:SFRLambdaAnti}
\end{figure}

To further unveil the connection between CGM gas spin and star formation and quenching activities at fixed stellar mass, we present Fig.~\ref{fig:CGMspinSFRAnticorrelation}, in which we show the distributions of the black hole accretion rate $\dot{M}_{\rm BH}/M_{\rm BH}$ and the specific star formation rate (sSFR), as a function of stellar mass $M_{\ast}$, for the three types of galaxies and their progenitors since $z=1$. Both quantities decrease with increasing galaxy masses. Galaxies are then divided into 8 bins according to their stellar masses. In each bin, galaxies are further split into two subgroups according to their CGM angular momenta $j_{\rm CGM}$, i.e., galaxies with $j_{\rm CGM}>j_{\rm CGM,50}$ (red) and with $j_{\rm CGM}<j_{\rm CGM,50}$ (blue), where $j_{\rm CGM,50}$ is the median $j_{\rm CGM}$ in each stellar mass bin. As can be seen, within the same mass range, galaxies with higher CGM angular momenta systematically exhibit lower star forming activity and lower feedback strength. 

It is worth noting that in Paper I we explicitly demonstrate that the cold-phase ($10^4\,{\rm K} < T < 2\times 10^4\, {\rm K}$) and hot-phase ($T > 10^5 \,{\rm K}$) CGM follow different kinematics: on average the cold-phase CGM possesses higher angular momentum and exhibits more tangential motion than its hot-phase counterparts. In Fig.\,\ref{fig:SFRLambdaAnti}, we present a clear anti-correlation between the sSFRs and the dimensionless spins (see Eq.~\ref{eq:lcgm} for definition) of the cold-phase CGM in star-forming galaxies (see also Figure 5 of Paper I). We note that we have left out the quenched galaxy sample in making the plot, as they have too little central gas contents and essentially too low SFRs.   

Such an anti-correlation strongly indicates a deep connection between the ambient gas angular momentum and the star-forming activity within a galaxy. Specifically, high CGM gas spins may prevent the outer gas from efficiently falling into galaxy centres, directly hampering a sustainable and efficient gas replenishment to fuel the central star formation as well as to feed the supermassive black holes. This may particularly have a larger impact on galaxies that happen to evolve into quiescent stages after having consumed a significant fraction of their central gas reservoirs: for a lack of sufficient gas supply (to the inner region), the galaxy could remain quenched. In this sense, this provides a complementary ``quenching'' mechanism, helping to keep a galaxy starved once it is quenched. Such a trend can be clearly seen in Section\,\ref{sec:GalDiffInCGMspin}, where we present the difference in CGM gas spins among different types of galaxies.

\section{From environmental angular momentum to CGM gas spin}
\label{sec:FromLSOAMtoCGM}

\begin{figure*}
\centering
\includegraphics[width=1.8\columnwidth]{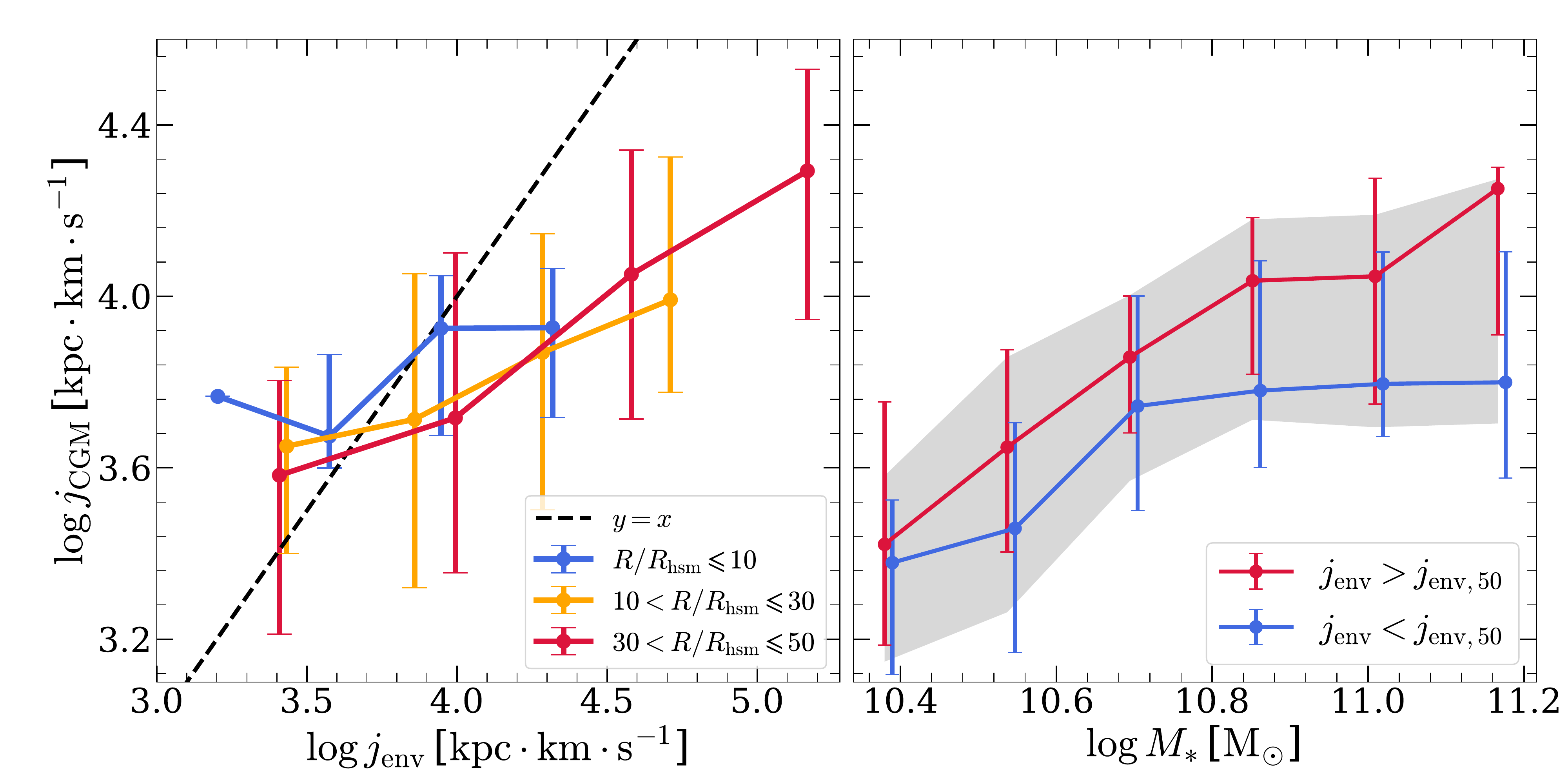}
\caption{Left: the correlation between the specific CGM spin $j_{\rm CGM}$ and the specific environmental angular momentum $j_{\rm env}$, for all three types of galaxies at $z=0$. The angular momenta are calculated in different radial ranges: $R/R_{\rm hsm}\,\leqslant\,10$ in blue, $10<R/R_{\rm hsm}\,\leqslant\,30$ in orange, and $30<R/R_{\rm hsm}\,\leqslant\,50$ in red, respectively. The error bars indicate the range from 16th to 84th percentiles ($\pm 1\sigma$). The black dashed line indicates $j_{\rm CGM}=j_{\rm env}$. Right: the distribution of the specific CGM angular momentum $j_{\rm CGM}$ as a function of galaxy stellar mass $M_{\ast}$ for all selected galaxies at $z=0$. Galaxies are divided into 6 stellar mass bins, with the grey shaded region indicating the $\pm 1\sigma$ region for all galaxies. In each bin, galaxies are further divided into two subgroups according to their environmental spin $j_{\rm env}$; i.e., galaxies with $j_{\rm env}>j_{\rm env,50}$ (red) and with $j_{\rm env}<j_{\rm env,50}$ (blue), where $j_{\rm env,50}$ is the median $j_{\rm env}$ in each bin. The error bars indicate the ranges from 16th to 84th percentiles ($\pm 1\sigma$) in individual subgroups.}
\label{fig:Jenv_Jcgm}
\end{figure*}

In Paper I of this series, we showed two example cases, where in-spiraling CGM gas was either associated with or triggered by interacting galaxies in the vicinity of the investigated galaxies (see Figures 7 and 8 therein). This already gives a hint that the large-scale environment may influence the CGM and its angular momentum via galaxy interactions. In the previous section, we demonstrate that the CGM gas spin would play an important role in regulating star formation and quenching activities inside a galaxy. In this section, we zoom out to a larger scale, and investigate specifically the connection between the CGM gas spin and the orbital angular momentum of neighbouring galaxies in the environment.

We first present, in the left panel of Fig.~\ref{fig:Jenv_Jcgm}, the specific CGM angular momentum $j_{\rm CGM}$ (see Eq.\,\ref{eq:j} for definition) versus the specific orbital angular momentum $j_{\rm env}$ from the neighbouring galaxies (see Eq.\,\ref{eq:jenv} for definition) for all three types of galaxies at $z=0$. The three colors show the relations evaluated within three different radial ranges. As can be seen, the CGM angular momentum shows a positive correlation with environmental angular momentum, already hinting at a deep connection between the two. In addition, within each (same) radial range, the averaged $j_{\rm env}$ is always systematically larger than $j_{\rm CGM}$, all the way out to $50R_{\rm hsm}$ ($\sim 250\,\rm kpc$). As shown in Fig.\,\ref{fig:GalComEnvAng} in Section\,\ref{sec:GalDiffInEnvAng}, the misalignment angles between the two angular momenta are moderately small. This essentially means that the overall modulation as created by the galaxy environment should be mainly in the form of injecting angular momentum to the CGM gas (see also \citealt{Stewart11CGMaccretion, Stewart17CWstreamHiAM}).

The right panel of the figure further illustrates the relation above but in a mass-independent manner. The panel shows the distribution of the specific CGM angular momentum $j_{\rm CGM}$ as a function of galaxy stellar mass $M_{\ast}$. For this distribution, galaxies are divided into 6 stellar mass bins, with the grey shaded region indicating the $\pm 1\sigma$ region for all galaxies. In each bin, galaxies are further divided into two subgroups according to their environmental spin $j_{\rm env}$; i.e., galaxies with $j_{\rm env}>j_{\rm env,50}$ (red) and with $j_{\rm env}<j_{\rm env,50}$ (blue), where $j_{\rm env,50}$ is the median $j_{\rm env}$ in each bin. As can be seen, galaxies (of similar masses) that live in higher angular momentum fields on average also have higher CGM angular momenta. We emphasize that this connection cannot start from inside-out, but may only happen as a consequence of the neighbouring galaxies transferring angular momentum to the CGM gas when they interact with and/or merge into the system. Given the interplay between the CGM motion and the star formation and feedback activities (as demonstrated in the previous section as well as in Paper I), this process essentially links the large-scale environment with the central star formation inside a galaxy, via modulating the CGM gas angular momentum.

\section{The fate of star-forming disks and quenched early-type galaxies}
\label{sec:FATES}

As demonstrated in previous sections, a galaxy's large-scale environment can affect star forming and quenching activities inside a galaxy, via modulation of the angular momentum of the CGM gas at intermediate distances. An implication could be that the states of star-forming disk and quenched early-type galaxies are simply consequences of different torque environments on larger scales. In Sections\,\ref{sec:GalDiffInCGMspin} and \ref{sec:GalDiffInEnvAng}, we present the differences in the ambient CGM and environmental angular momenta, respectively; and in Section\,\ref{sec:GalDiffInEvolution} we show how this is reflected in the histories of three different types of galaxies.  

\subsection{Difference in CGM spins}
\label{sec:GalDiffInCGMspin}

\begin{figure*}
\includegraphics[width=0.9\columnwidth]{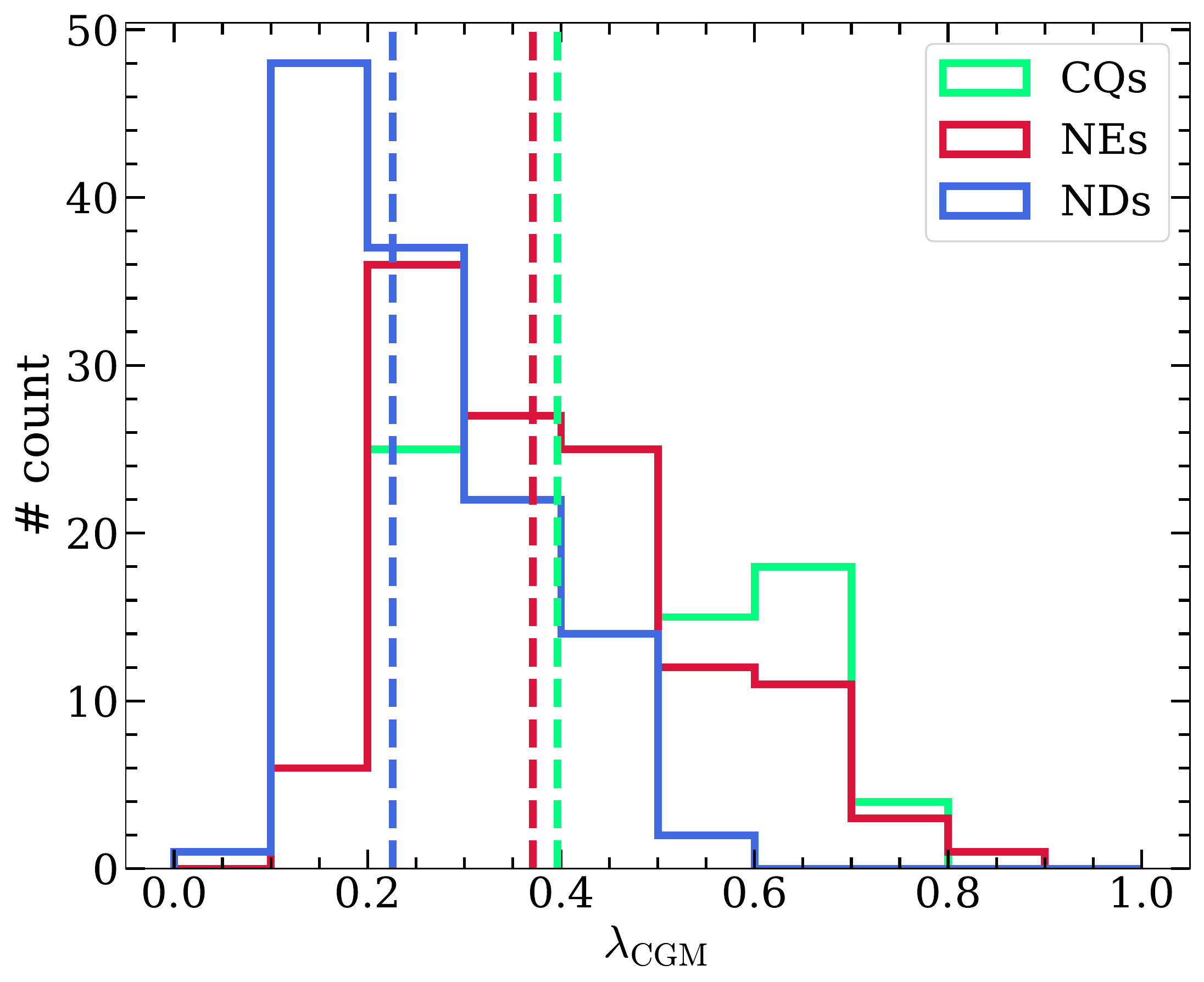}
\includegraphics[width=0.92\columnwidth]{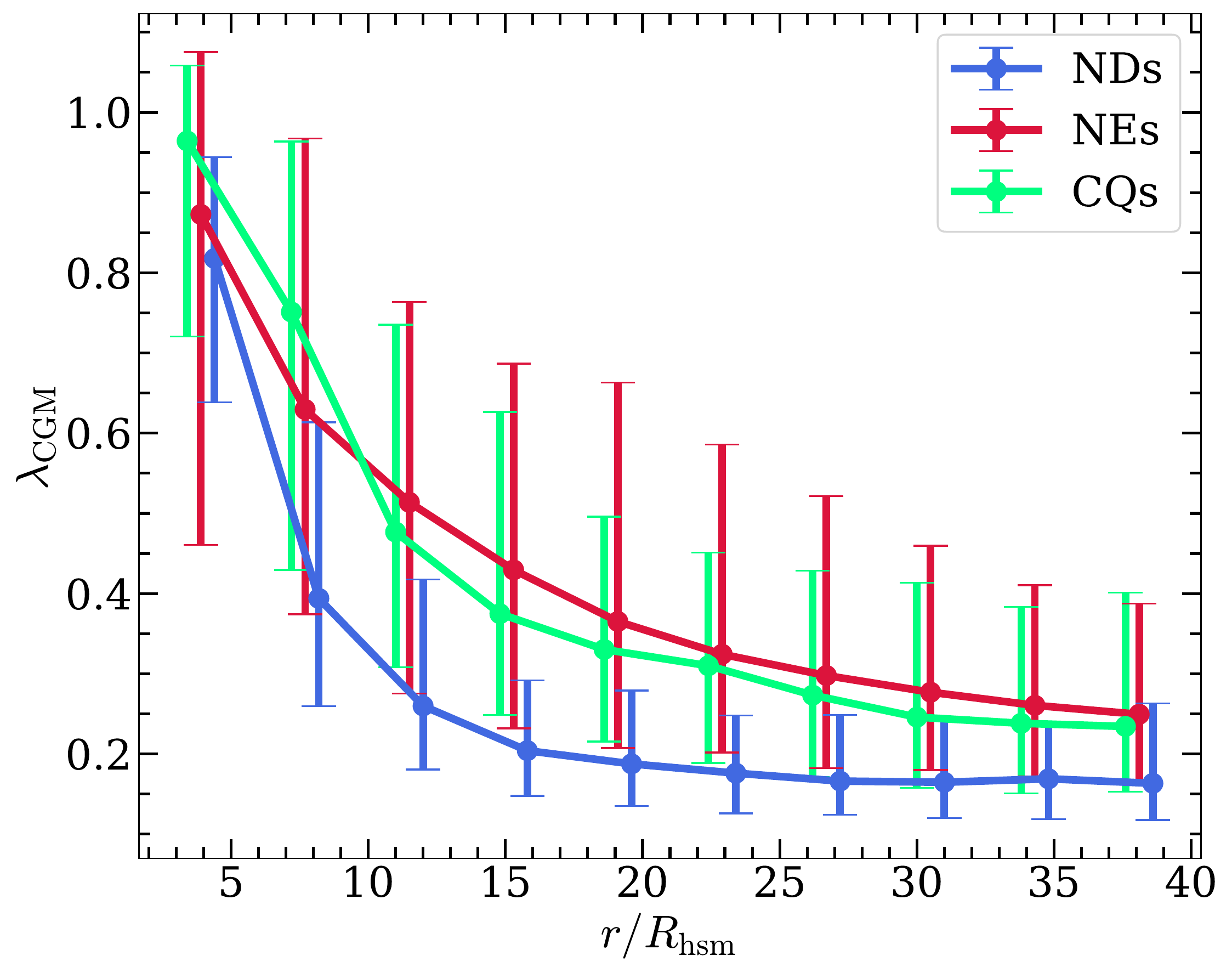}
\caption{Left: histograms of the CGM dimensionless spin $\lambda_{\rm CGM}$ (evaluated within a radial range of $30-300$ kpc, see Eq.\,\ref{eq:lcgm} for definition) for three types of galaxies at $z=0$. The dashed line indicates the median value in each case. Right: radial profiles of the CGM dimensionless spin $\lambda_{\rm CGM}$ for the same galaxy samples. The error bars indicate the ranges from 16th to 84th percentiles ($\pm 1\sigma$) in individual cases.}
\label{fig:lambda_galaxies}
\end{figure*}

\begin{figure*}
\includegraphics[width=2\columnwidth]{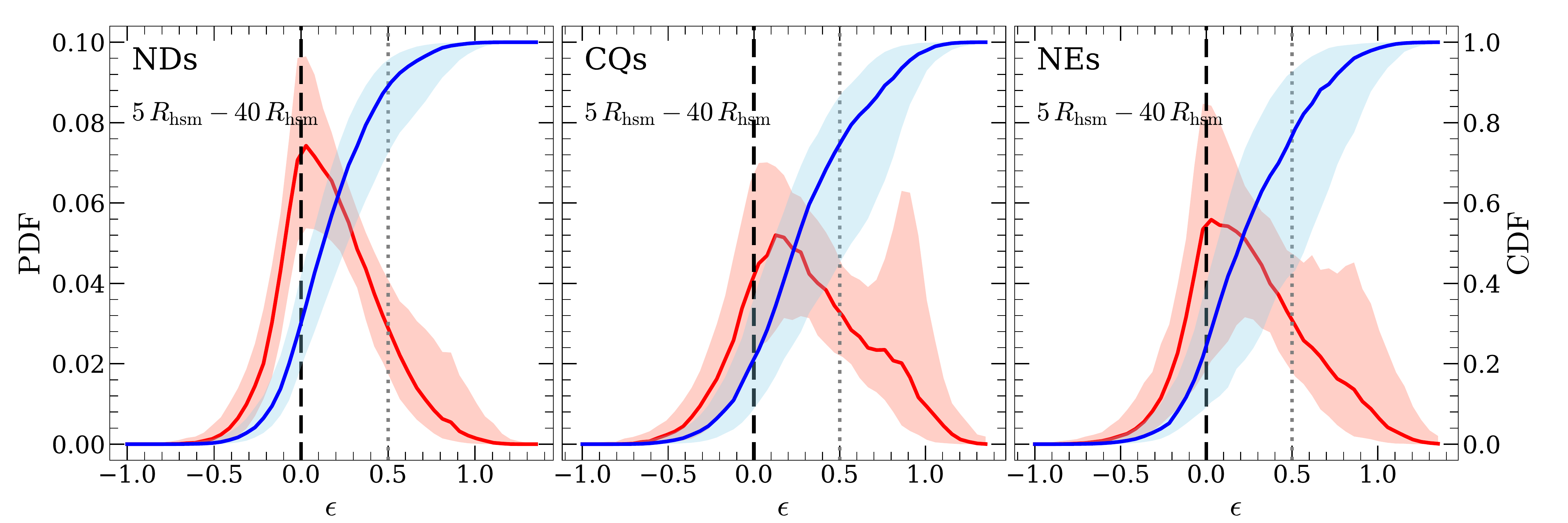}
\caption{Circularity ($\epsilon$) distributions of the CGM gas within $5R_{\rm hsm}<r<40R_{\rm hsm}$ for NDs (left), CQs (middle), and NEs (right). In each panel, the mass probability distribution function (PDF) of $\epsilon$ is shown by the red curve (corresponding to the left Y-axis) and the cumulative mass distribution function (CDF) by the blue curve (corresponding to the right Y-axis), with the shaded regions indicating the ranges from 16th to 84th percentiles ($\pm 1\sigma$). The black dashed line indicates $\epsilon=0$ and the grey dotted line indicates $\epsilon=0.5$, above which the orbits of the gas cells are close to circular orbits.}
\label{fig:circularity}
\end{figure*}

As demonstrated in Section\,\ref{sec:CGMvsSFR}, galaxies with higher CGM gas spins tend to have less-active star formation and feedback activities. Then would it be true that present-day quenched galaxies have higher CGM (dimensionless) spins than their star-forming disk counterparts? To answer this question, we present the left panel of Fig.\,\ref{fig:lambda_galaxies}, where the distributions of the dimensionless CGM spin $\lambda_{\rm CGM}$ (see Eq.~\ref{eq:lcgm}) are plotted for three types of galaxies at $z=0$. As expected, the CGM spins of quenched early-type galaxies are systematically higher than those of star-forming disk galaxies.

We shall emphasize that the statement of the CGM gas possessing high angular momentum does {\it not} mean that the system becomes {\it globally} rotationally supported at larger radii. In fact, as the radius increases, the CGM gas becomes less and less rotationally supported, which can be seen in the right panel of Fig.\,\ref{fig:lambda_galaxies}, where the detailed radial profiles of the CGM dimensionless spin $\lambda_{\rm CGM}$ are presented. The cold circumgalactic gas, which dominates the CGM angular-momentum budget (see Figure 6 of Paper I), carries out rather localized in-spiral motions, responsible for possessing high angular momenta. Again, the difference among the three galaxy types clearly shows up especially at larger distances. 

To better demonstrate the above-presented differences in CGM gas spins among different types of galaxies, we further present Fig.\,\ref{fig:circularity}, where the circularity ($\epsilon$) distributions of the CGM gas within $5R_{\rm hsm}<r<40R_{\rm hsm}$ for three types of galaxies are presented. We first notice that the mass fractions of the CGM gas being retrograde (with respect to the total CGM spin) are relatively low: only $5\%-6\%$ have $\epsilon < -0.25$. In addition, the mass fractions of the CGM gas having $\epsilon > 0.5$ (i.e., having high angular momenta) in present-day quenched galaxies (i.e. CQs and NEs) on average are by a factor of 2 higher than that in present-day star-forming disk galaxies (24.9\% for NEs, 26.7\% for CQs, and 11.9\% for NDs).
%again supporting the idea that the gas in quenched galaxies (NEs and CQs) is in general on more stable (higher angular momentum) orbits and can not easily fall into the galaxies.}

All the findings above strongly suggest that high CGM angular momentum may provide a complementary explanation for galaxies {\it remaining} quenched (once they are quenched). This can also explain the fact that some of the quenched early-type galaxies are seen to have in-spiraling cold CGM features at larger radii. In particular in the case of fast-rotators, ring-like HI morphologies may often be present at the galaxy outskirts, indicating high angular momentum (e.g., \citealt{Oosterloo_et_al.(2007),Christlein_et_al.(2008),Sancisi_et_al.(2008),Putman_et_al.(2009),Lemonias_et_al.(2021)}; also see figure 11 of \citealt{Lu_et_al.(2021b)}). We note that absorption line studies on CGM gas kinematics in star-forming and quenched galaxies would provide valuable observation evidence to test this scenario. We would like to in particular point out two observations of such from \citet{Huang16AbsorptionLRG,Huang21MgIIAbsorption}. It is truly intriguing that the dispersion of the measured ratios between the line-of-sight velocities and the escape velocities of the CGM gas around more massive passive galaxies is lower than those of star-forming galaxies. This is not seen for our galaxy samples. We suspect mass-dependent systematics could be at work to cause the observed pattern and we call for more studies in this regard.

\subsection{Difference in environmental angular momenta}
\label{sec:GalDiffInEnvAng}

\begin{figure*}
\includegraphics[width=1.8\columnwidth]{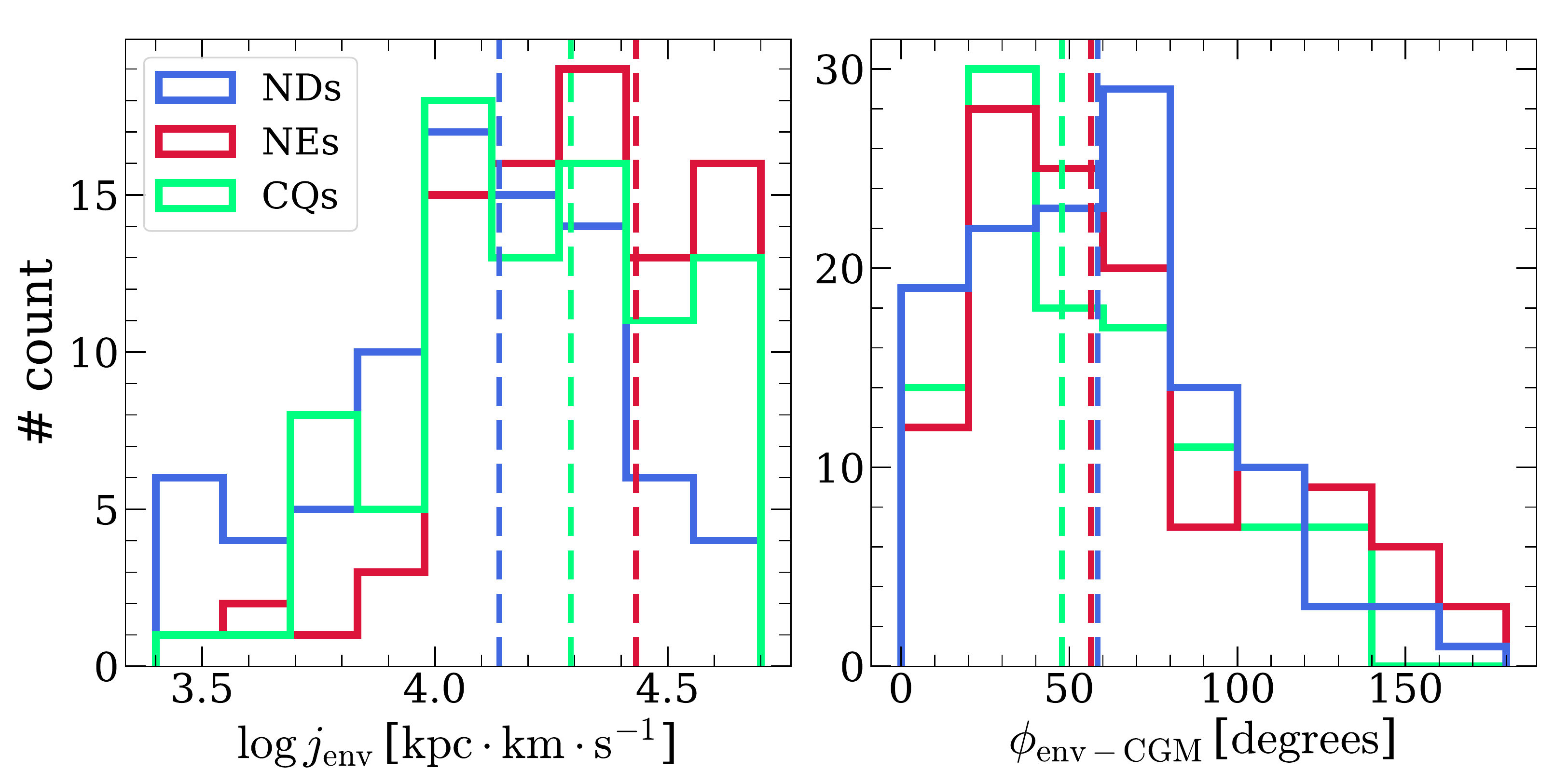}
\caption{Histograms of the specific environmental angular momentum $j_{\mathrm{env}}$ (left panel; evaluated in the distance range of $50-300$ kpc from the host galaxy; see Eq.~\ref{eq:jenv} for definition) and of misalignment angles between $\textbf{j}_{\rm CGM}$ and $\textbf{j}_{\mathrm{env}}$ (right panel), for NDs (blue), CQs (green), and NEs (red) selected at $z=0$. In each panel, the dashed lines indicate the median values of the investigated parameters for the three types of galaxies.}
\label{fig:GalComEnvAng}
\end{figure*}

As demonstrated in Section\,\ref{sec:FromLSOAMtoCGM}, galaxies which reside in an environment that induces larger orbital angular momentum from neighbouring galaxies also have higher CGM angular momentum. In the previous subsection, it is shown that present-day quenched early-type galaxies on average have higher CGM spins than their star-forming disk counterparts. Then would it be true that the former also reside in higher angular momentum environments than the latter?

To answer this question, we present Fig.\,\ref{fig:GalComEnvAng}, which shows the distributions of: \textbf{(1)} the specific environmental angular momentum $j_{\mathrm{env}}$ (left panel; evaluated in the distance range of $50-300$ kpc from the host galaxy; see Eq.~\ref{eq:jenv} for definition) and \textbf{(2)} the misalignment angle $\phi_{\rm env-CGM}$ between the two angular momentum vectors (right panel). As expected, the large-scale environment around quenched early-type galaxies indeed on average possesses higher orbital angular momenta than that around star-forming disks. Interestingly, the misalignment angle $\phi_{\rm env-CGM}$ shows rather similar distributions among the three types of galaxies, all peaking at lower angles (with a median of $\sim 50^{\circ}$), again suggesting an inherent connection between their angular momenta.

\subsection{Difference in evolution histories}
\label{sec:GalDiffInEvolution}

\begin{figure*}
\includegraphics[width=1\columnwidth]{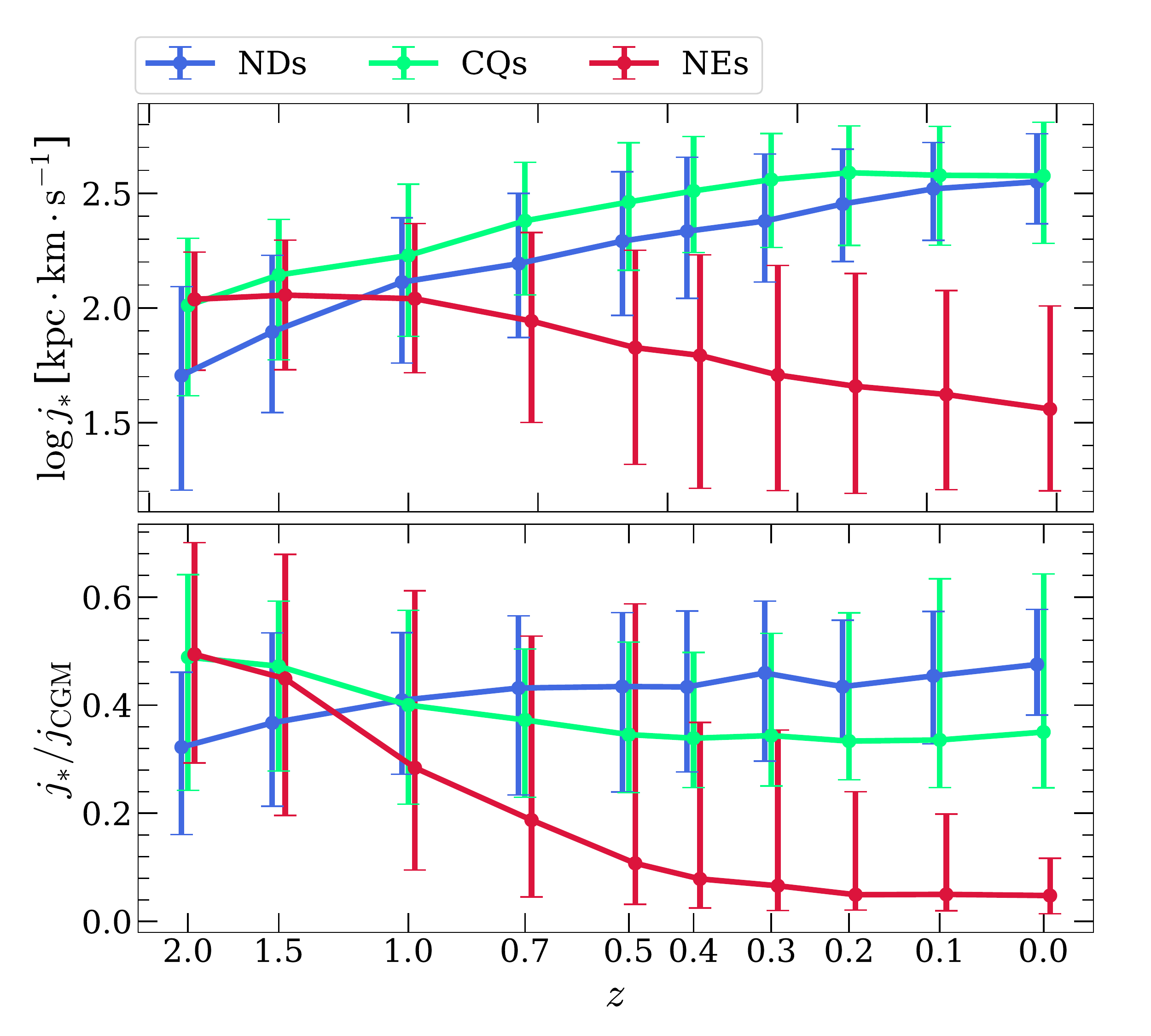}
\includegraphics[width=1\columnwidth]{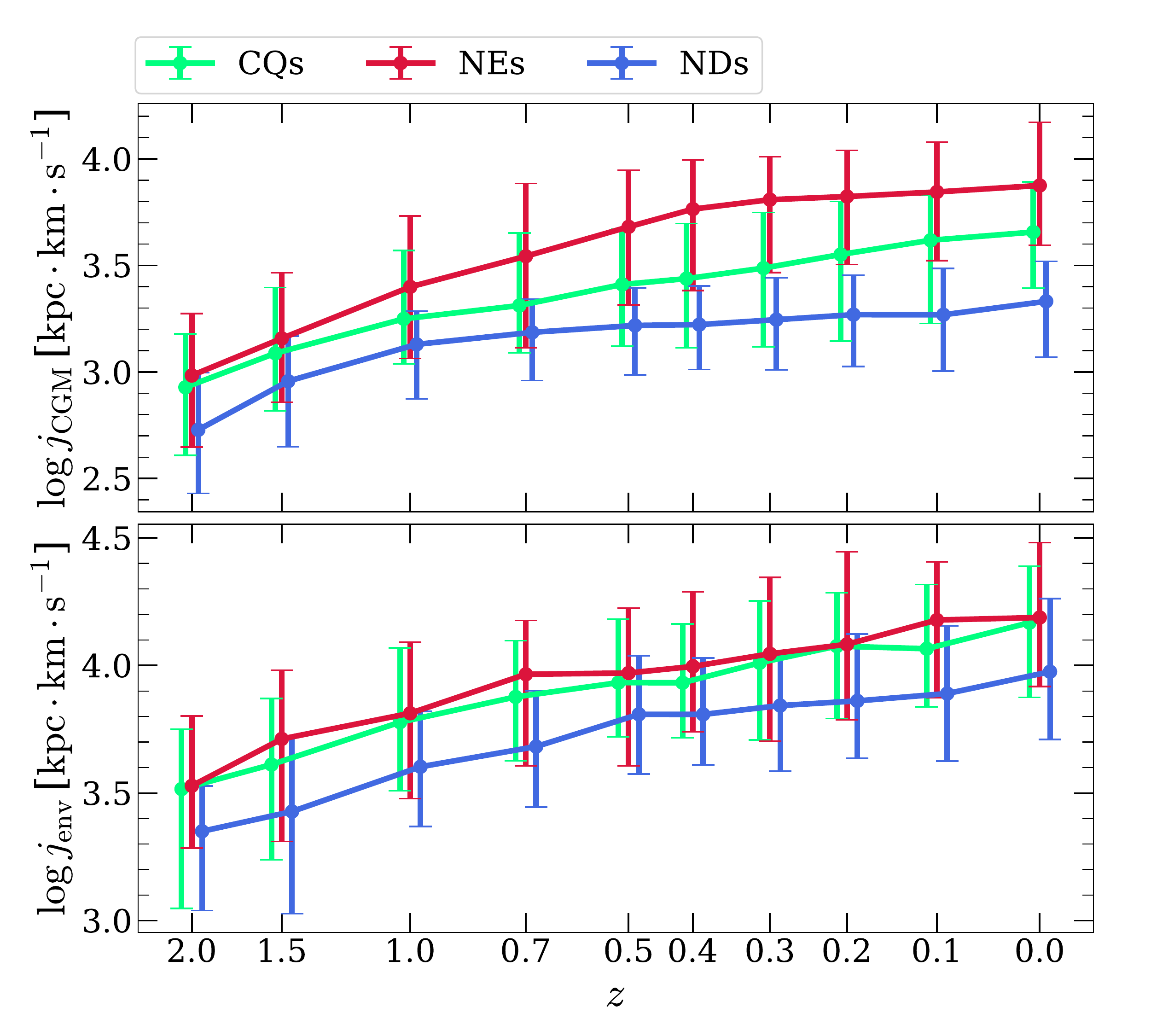}
\caption{The redshift evolution of: \textbf{(1)} the stellar spin within $2R_{\rm hsm}$ ($\log\,j_{\ast}$, top left), \textbf{(2)} the specific CGM gas angular momentum ($\log\,j_{\rm CGM}$, top right; evaluated at $r>2R_{\rm hsm}$, see Eq.\,\ref{eq:j} for definition), \textbf{(3)} the ratio between the two angular momenta above (i.e., $j_{\ast}/j_{\rm CGM}$, bottom left), and \textbf{(4)} the specific environmental angular momentum ($\log\,j_{\rm env}$, bottom right; see Eq.\,\ref{eq:jenv} for definition). In each panel, CQs, NEs, and NDs are indicated by green, red, and blue lines, with error bars indicating the range from the 16th to the 84th percentiles ($\pm 1\sigma$).}
\label{fig:jcgm_jenv_evolve}
\end{figure*}

As already demonstrated in the previous two subsections, differences are clearly present in the CGM gas spins and the environmental angular momenta among three different types of galaxies. Then the next question is how long ago have such ambient and environmental differences been in place?

In order to answer this question, we present the redshift evolution of the stellar disk spin, the CGM gas spin and the environmental orbital angular momentum since $z=2$ in Fig.~\ref{fig:jcgm_jenv_evolve}. As can be seen from the left panels, present-day quenched early-type galaxies used to have significantly higher stellar angular momenta at early times, during which their disk morphologies and kinematics were well established in comparison to their star-forming disk counterparts. With time, fast-rotators (slow-rotators) steadily gain (lose) stellar angular momentum, as a consequence of prograde (retrograde) mergers (see \citealt{Lu_et_al.(2021b)}). The {\it specific} stellar angular momenta of present-day star-forming disc galaxies have increased by a factor of $\sim 2.7$ from $z=1$ to $z=0$ and by a factor of $\sim 7$ from $z=2$ to $z=0$. This is well consistent with the result of \citet{Swinbank_et_al.(2017)} from observations (a factor of $\sim 3$ from $z=1$ to $z=0$), and that reported by \citet{Peng_et_al.(2020)} and \citet{Renzini(2020)} through the scaling relations (a factor of $\sim 7$ from $z=2$ to $z=0$). 
%It implies that the growth of the stellar disc angular momentum is not the simple combination of the growths of galaxy mass and size.
As a result, present-day fast-rotating early-type galaxies and star-forming disks have much higher stellar angular momenta than slow-rotating elliptical galaxies. 
Zooming out to larger scales (right panels), slow-rotators have always had the highest CGM spins and environmental angular momenta since at least $z\sim 2$, while star-forming disk galaxies have always had the lowest ambient and environmental angular momenta. 

Such differences in larger-scale angular momentum among different types of galaxies existing consistently across a wide range of redshifts suggests that the different fates of galaxies may originate from an early epoch. Specifically, the initial conditions of the large-scale torque fields that are fixed at the birth sites, to a certain degree, have already determined the coarse-grain ``paths'' of later-stage galaxy interactions, through which the CGM gas is modulated by inheriting angular momentum from the torque environment.

\section{Observational Predictions}
\label{sec:observation}

\subsection{Coherent kinematics from stacked velocity maps}

\begin{figure}
\centering
\includegraphics[width=1\columnwidth]{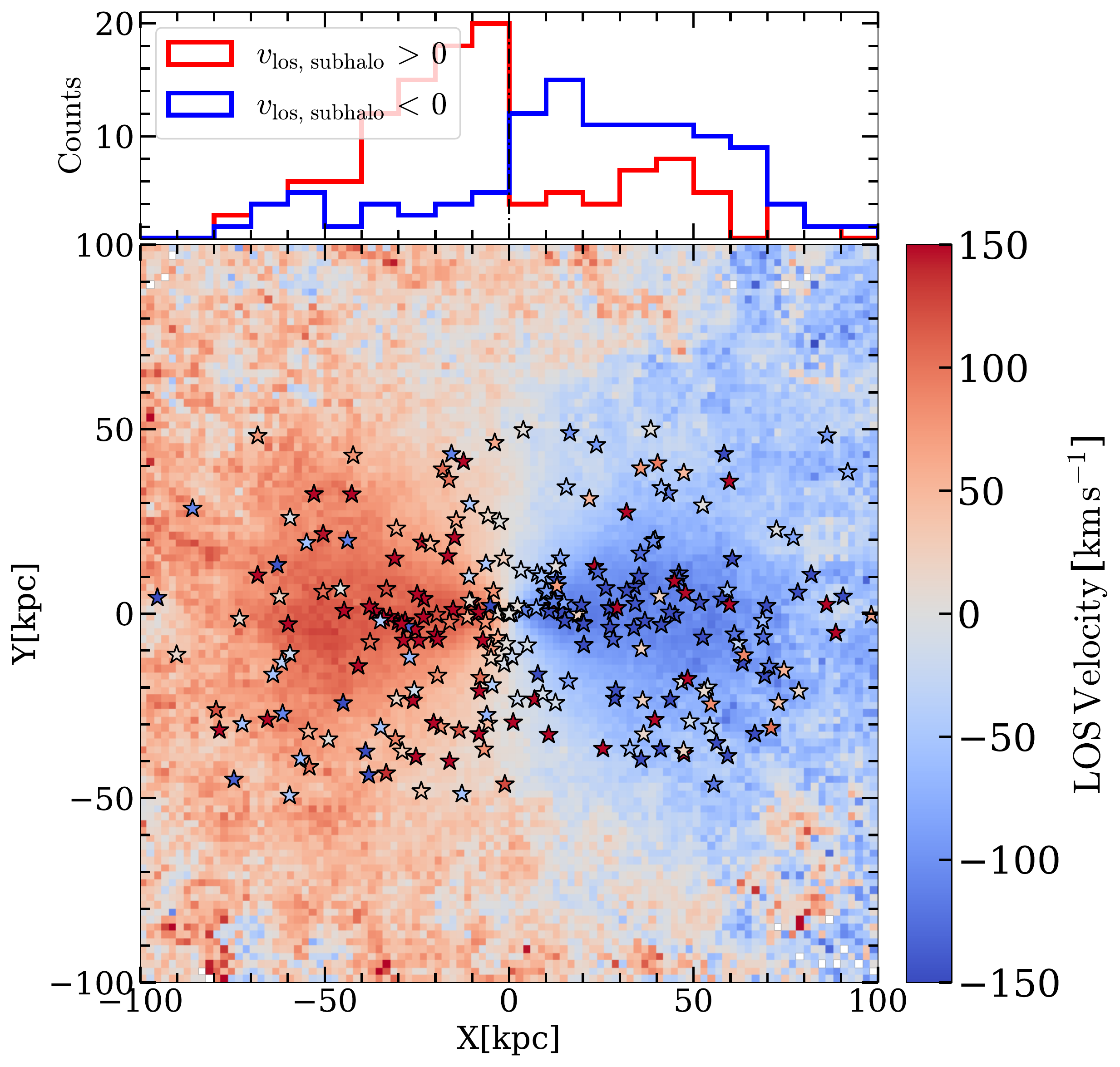}
\caption{Stacked line-of-sight velocity fields of all the star-forming disks and the quenched but dynamically cold early-type galaxies at $z=0$. Before stacking, all the galaxies are rotated to edge-on views according to their stellar disks such that the galaxy major axes are along the $X$-axis in the figure. A further rotation with a random disk inclination angle between $-60^{\circ}$ and $60^{\circ}$ along the major axis is then applied to each galaxy, in order to mimic a nearly edge-on view from an observation. 
The color-coded maps present the stacked signal of the line-of-sight velocity fields of the cold ($10^4\,\mathrm{K} < T<2\times 10^4\,\mathrm{K}$) CGM gas. The star symbols with colors indicate the line-of-sight velocities of galaxies (with projected $|Y|<50\,\rm kpc$) in the vicinity of the hosts. The histograms on the top of the figure show the spatial distributions along the $X$-axis (the major axis of the stellar disk) for neighbouring galaxies with positive (red-shifted) and negative (blue-shifted) line-of-sight velocities.} 
\label{fig:stackedLOSV}
\end{figure}

The inherent connection between the environmental angular momentum and the CGM gas spin, as demonstrated in Section\,\ref{sec:FromLSOAMtoCGM}, would become even clearer when we over-plot \textbf{(1)} the stacked line-of-sight velocities of neighbouring galaxies in the vicinity of the dynamically cold galaxies, on top of \textbf{(2)} the stacked line-of-sight velocity fields of the CGM gas. This is shown in Fig.\,\ref{fig:stackedLOSV}, where the nearly edge-on views of all the star-forming disk galaxies and fast-rotating early-type galaxies are stacked according to the stellar disk orientations such that the apparent major axes of the stellar disks are lined up and the red-shifted sides of the velocity fields are put on the same side in the sky plane. We note that to make Fig.\,\ref{fig:stackedLOSV}, we have not included slower-rotating elliptical galaxies. This is simply because they do not have well-established rotating stellar disks to line up the frames for the purpose of stacking. However, these systems are {\it not} lacking of in-spiralling cold CGM gas (e.g., see Figure 8 of Paper I).

It is truly intriguing to see such a coherent rotation pattern systematically existing among the stellar disk (a few kpc), the CGM gas (tens of kpc) and neighbouring galaxies (up to $\sim$100 kpc). We also note that such coherent kinematics from stacked galaxy populations not only exists at $z=0$ but also shows up for dynamically cold galaxies up to $z\sim 1$. 
%\textcolor{red}{\citet{Lu_et_al.(2021b)} have reported that NEs show higher fraction of environment members being retrograde with respect to their stellar spins, compared to their dynamically cold counterparts (i.e. NDs and CQs). This also means that the misalignment angle between the spin axes of the stellar disc and the CGM gas is larger in NEs than that in CQs and NDs.}
It is worth noting that on an individual basis, median misalignment angles of $40^{\circ}-50^{\circ}$ are typically seen between the stellar disks and the CGM gas in star-forming disks and fast-rotating early-type galaxies\footnote{In comparison, slow-rotating early-type galaxies on average have higher misalignment angles between their stellar disks and the CGM gas. This has largely to do with the fact that a higher fraction of merging satellite galaxies being retrograde with respect to the stellar spin of the host (\citealt{Lu_et_al.(2021b)}).}. When stacked together along the stellar disks, however, the misalignment in individual galaxies cancels out, only leaving behind a cone-shaped co-rotation feature between the CGM and stellar disks in dynamically cold galaxies, as is shown in Fig.\,\ref{fig:stackedLOSV}.

Observationally, co-rotations between the CGM and the central stellar disks have been detected in many recent studies (e.g., \citealt{Danovich15CWstreamToDisk, Ho_et_al_2017, Martin_et_al_2019, Zabl_et_al_2019}; see \citealt{Tumlinson_et_al.(2017)} and references therein). However observations of co-rotation between the CGM gas and galaxies in close neighbourhoods have been scarce and tentative. Among the three types of kinematic measurements, line-of-sight velocities of the cold CGM gas around a fairly large sample of galaxies would be the most tricky one to obtain. However, modern observations make the measurement of CGM kinematics possible. This can be achieved through dedicated absorption line observations towards quasar sight lines (e.g., \citealt{Bouche_et_al.(2017),Stewart(2017), Martin_et_al_2019, Wilde_et_al.(2021)}) or through recent emission studies (e.g., \citealt{Cai_et_al.(2017), Cai_et_al.(2019), Arrigoni_Battaia_et_al.(2019)}). For these galaxies, resolved stellar kinematic maps or resolved stellar disk morphologies would also be needed in order to stack galaxies along the same rotation/disk directions. At larger distances, the line-of-sight velocity field of neighbouring galaxies can be obtained through spectroscopic measurements which can resolve relative line-of-sight velocities below 100 km/s. Such observations at multiple scales would provide best support for the influence of angular momentum from $\sim 100$ kpc all the way down to the stellar disk scale inside a galaxy.

\subsection{Anti-correlation between environmental angular momenta and galaxy star formation rates}

\begin{figure*}
\includegraphics[width=2\columnwidth]{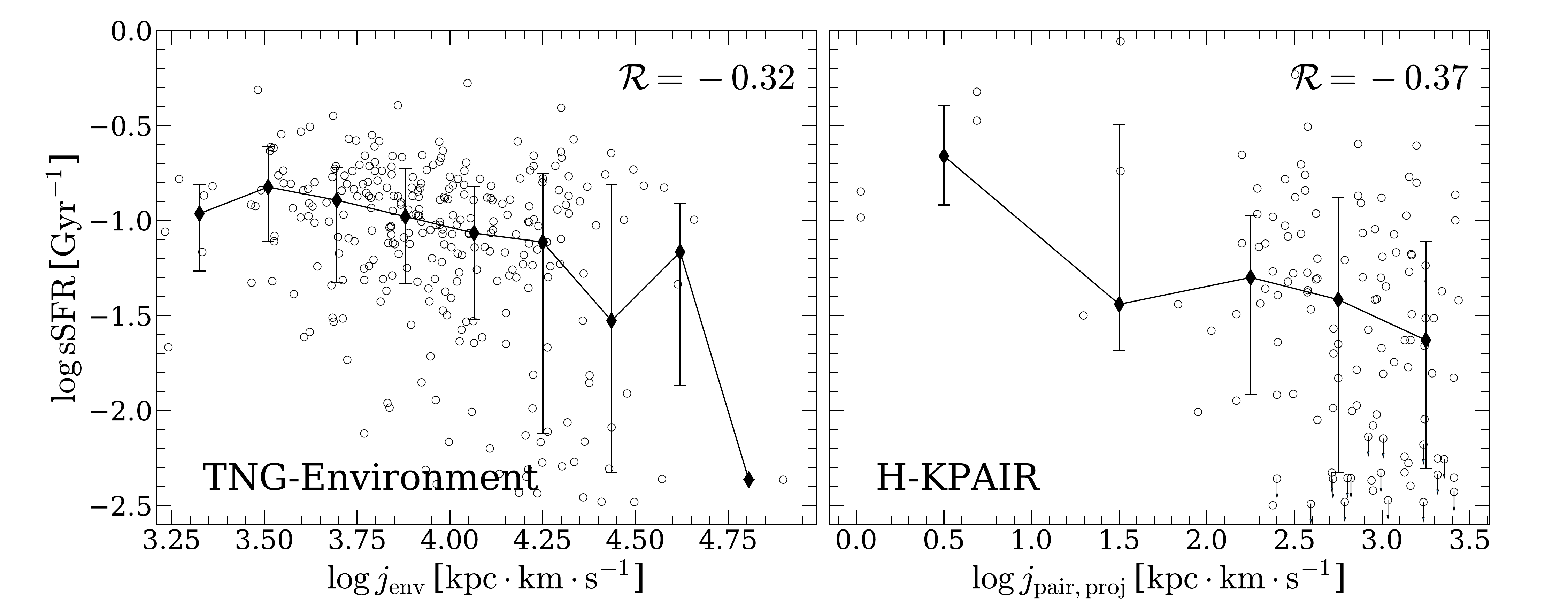}
\caption{Left: correlation between the specific star-formation rate ($\log\,\mathrm{sSFR}$) and the environmental angular momentum ($\log\,j_{\rm env}$, see Eq.\,\ref{eq:jenv} and Section \ref{sec:def} for definition) of the TNG galaxy samples at $z\leqslant 0.1$ (the same redshift range as the H-KPAIR sample on the right). The median profile of $\log\,\mathrm{sSFR}$ as a function of $\log\,j_{\rm env}$ is indicated by the black lines. Right: correlation between the specific star-formation rate ($\log\,\mathrm{sSFR}$) and the angular momenta of galaxy pairs ($\log\,j_{\rm pair,proj}$, see Eq.~\ref{eq:jpair} for definition) in the H-KPAIR sample. The error bars in both panels indicate the ranges from 16th to 84th percentiles ($\pm 1\sigma$). $\mathcal{R}$ is the Pearson correlation coefficient for the two investigated parameters.}
\label{fig:jpair_ssfr}
\end{figure*}

Regarding the impact of ambient and environmental angular momenta on central star formation, we specifically propose to seek for an anti-correlation between galaxy star formation rates and specific environmental angular momenta. As an example, we present, in the left panel of Fig.\,\ref{fig:jpair_ssfr}, the expected anti-correlation between $\log\,\mathrm{sSFR}$ and $\log\,j_{\rm env}$ (see Eq.\,\ref{eq:jenv} and Section \ref{sec:def} for definition) for the TNG galaxy samples and at $z\leqslant 0.1$. To characterize the relation between $\log\,\mathrm{sSFR}$ and $\log\,j_{\rm env}$, we simply employ the Pearson correlation coefficient $\mathcal{R}$. For the TNG galaxy samples, $\mathcal{R} \sim -0.32$ indicates a moderate anti-correlation therein. In practice, the 3D phase space information would not be fully available in order to evaluate $j_{\rm env}$. However, we note that in practice it is possible for the environment angular momentum to be roughly estimated using {\it projected} distances and relative {\it line-of-sight} velocities of neighbouring galaxies. In particular, we expect such an approximation to work better in close galaxy pairs with larger mass ratios (corresponding to smaller mass gaps between the two galaxies of a pair; see also \citealt{Patton20InteractingDistanceBoostingSFR}).

As a pilot study, we take 123 galaxies with $\log\,\mathrm{sSFR}/\mathrm{Gyr^{-1}}>-2.5$ from 88 close ``major-merger'' pairs in the H-KPAIR sample \citep{2016ApJS..222...16C}, which is a sub-sample of a complete and unbiased $Ks$-band (of the Two Micron All Sky Survey) selected galaxy pair sample - KPAIR \citep{2009ApJ...695.1559D}. All of the paired galaxies have spectroscopic redshifts $z < 0.1$ (but excluding those with recession velocities less than 2000\,km/s). Their stellar masses $M_{\ast}$ ranging from $10^{9.8} \mathrm{M_{\odot}}$ to $10^{11.4} \mathrm{M_{\odot}}$ were estimated using the $Ks$-band magnitudes and the SFRs were calculated based on the FIR-sub-millimeter spectral energy distribution fitting carried out using the HERSCHEL data. The galaxy pairs were required to have projected separations $5h^{-1}{\rm kpc} < s_{\rm proj} < 20h^{-1}{\rm kpc}$ and relative line-of-sight velocities $\Delta v_{\rm los}<500$\,km/s. The $Ks$-band magnitude difference between the two galaxies in each pair was required to be within 1 mag, corresponding to a (secondary-to-primary) mass ratio larger than 0.4, in order to guarantee a sufficiently small mass difference between the two galaxies (see \citealt{2016ApJS..222...16C} for details). 

In the right-hand side panel of Fig.\,\ref{fig:jpair_ssfr}, we present the distribution of $\rm \log sSFR$ as a function of $\log j_{\rm pair,\,proj}$ - the specific orbital angular momentum of the pair (with respect to the centre of mass), which is estimated as:
\begin{equation}
\label{eq:jpair}
\log j_{\rm pair,proj}  = \log \, \left[ s_{\rm proj} \Delta v_{\rm los} \frac{M_{\ast,1} M_{\ast,2}}{(M_{\ast,1} + M_{\ast,2})^2} \right],
\end{equation}
where subscripts 1 and 2 denote the two galaxies in the pair. As can be seen, a marked anti-correlation is, as expected, present among the data with the Pearson correlation coefficient being $\sim -0.37$, similar to that of TNG samples. Here we call for bigger samples of galaxy pairs in order to provide further observational support for the proposed anti-correlation between environmental angular momenta and central star formation activity.

\section{Conclusions and Discussion}
\label{sec:conclusion}

In this study, we use the IllustrisTNG-100 simulation to investigate the connections among the large-scale environment, the CGM spin and the central star-forming and quenching activities. We find that at larger distances, the CGM spins positively correlate with the environmental angular momenta due to the orbital motions of neighbouring galaxies (Fig.\,\ref{fig:Jenv_Jcgm}), whose velocity fields show coherent rotations with respect to that of the CGM gas, which further co-rotate with galaxy stellar disks on smaller scales (Fig.\,\ref{fig:stackedLOSV}). These connections support a hierarchical inheritance of angular momentum from the large-scale environment all the way to a galaxy's star-forming disk (see Section~\ref{sec:FromLSOAMtoCGM}). At smaller radii, the star formation rates and the supermassive black hole accretion rates, all anti-correlate with the CGM spins: less actively star-forming galaxies systematically exhibit higher CGM spins than their actively star-forming counterparts within the same stellar mass range (Figs.\,\ref{fig:CGMspinSFRAnticorrelation} and \ref{fig:SFRLambdaAnti}). This can be readily understood if higher CGM angular momentum prevents the gas from efficiently infalling and feeding the central star-forming gas reservoirs, resulting in less active star formation (see Section~\ref{sec:CGMvsSF}). Such a connection can even serve as a mechanism that can keep a galaxy quenched once it is quenched. The above-presented co-rotation motion and anti-correlation relations can be tested by observing the orbital properties of group galaxies or galaxy pairs through spectroscopic observations and by measuring the CGM velocity fields through absorption line observations towards quasar sight lines (see Figs.\,\ref{fig:stackedLOSV} and \ref{fig:jpair_ssfr} and Section~\ref{sec:observation}).

The large-scale environment plays an important role in affecting the evolution of a galaxy. It modulates the CGM by injecting angular momentum to it through galaxy interactions (e.g., mergers and fly-bys). Such an angular momentum modulation upon the CGM
has a significant impact on the galaxy's fate in terms of its star formation and quenching activities (see Section~\ref{sec:FATES}). Specifically, we find that present-day quenched early-type galaxies have always had much higher ambient and environmental angular momenta since as early as $z \sim 2$, in contrast with their star-forming disk counterparts (Figs.\,\ref{fig:lambda_galaxies}, \ref{fig:circularity}, \ref{fig:GalComEnvAng}, and \ref{fig:jcgm_jenv_evolve}). In line with this, present-day star-forming disk galaxies have always had sufficient feeding to the central gas reservoirs from larger distances. This can rejuvenate a galaxy in its quenched phase and bring it back on track to the star forming sequence. The galaxy thus would go through episodic star formation, intermittent with quenched phases (as demonstrated in Paper I). Present-day quenched early-type galaxies, on the other hand, had gone through early disk assemblies through rapid star formation (also see \citealt{Lu_et_al.(2021b)}). Their central gas reservoirs had been quickly consumed due to starbursts and AGN activities triggered by major mergers (e.g., \citealt{Barnes_et_al.(1992),Barnes_et_al.(1991),Barnes_et_al.(1996),Mihos_et_al.(1996), Barnes_et_al.(2002),Naab_et_al.(2003),Bournaud_et_al.(2005),Bournaud_et_al.(2007),Johansson_et_al.(2009a),Tacchella_et_al.(2016),Rodriguez-Gomez_et_al.(2017)}). At some point, as a cycle of star formation and feedback (which can effectively keep the cold gas stay outside the central region because of pressure, see Figure 6 of Paper I) dies down, these galaxies then transition into a quenched phase. However, acquiring overly high angular momenta from the large-scale environment prevents the cold CGM from efficiently flowing in to further feed the central gas reservoirs. These galaxies may remain quenched ever since. Therefore the large-scale environment can initiate a mechanism (referred to as ``angular momentum quenching'', see Section\,\ref{sec:AMQ}) that is directly executed by the CGM, which starves a galaxy and keeps it quenched once it is quenched. Below we present some further discussion on a number of key issues related to this work.

\subsection{The quenched state}

\begin{figure}
\includegraphics[width=1\columnwidth]{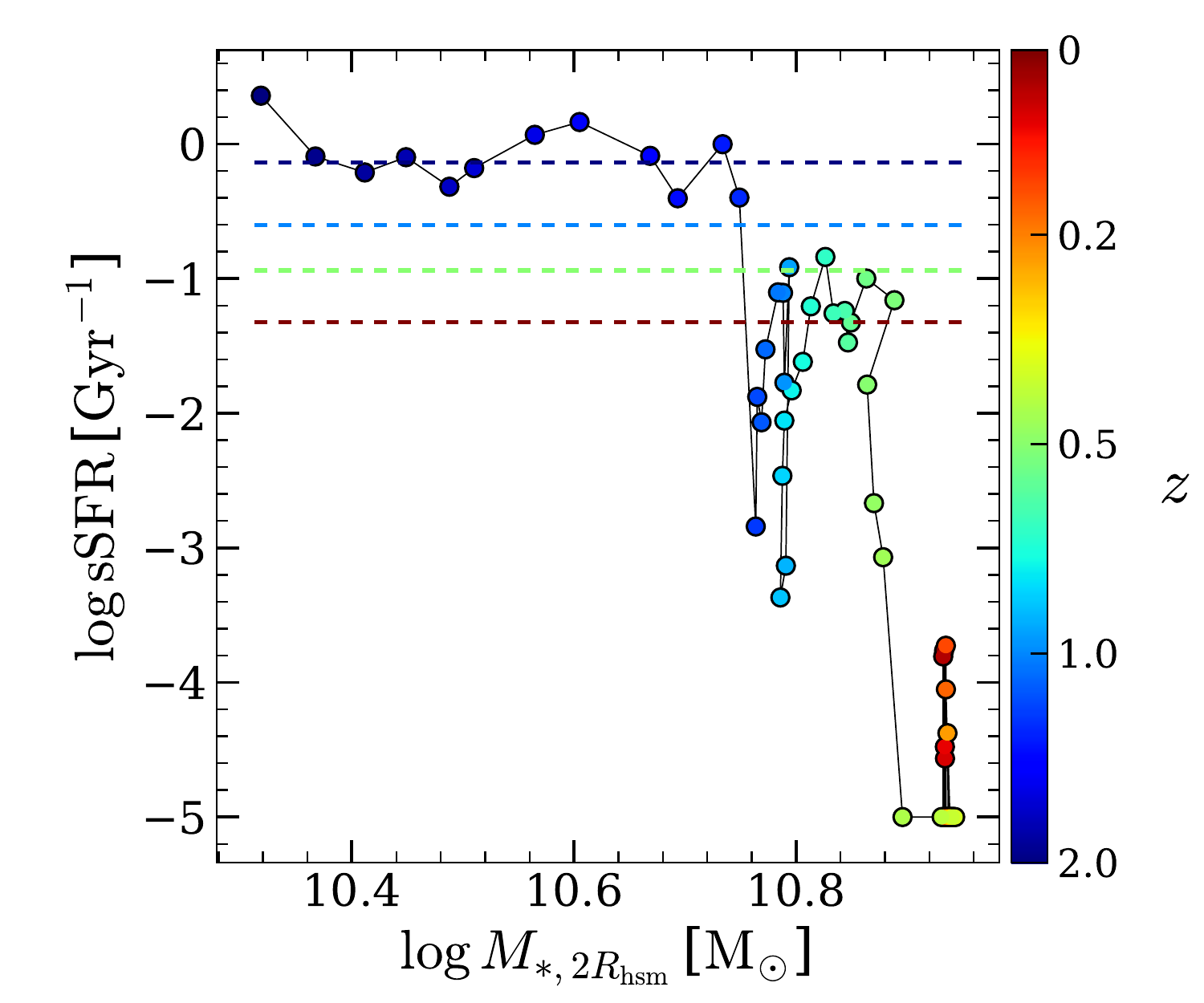}
\caption{The redshift evolution in the $\log\,\mathrm{sSFR}-\log\,M_{\ast}$ plane for a present-day quenched elliptical galaxy (ID: 407204). Four dashed lines from top to bottom indicate the lower boundary in $\log\,\mathrm{sSFR}$ for the main sequence galaxies at $z=2,\,1,\,0.5,\,0$ (indicated by colors), below which galaxies are regarded as quenched. The lower boundary is calculated as the 16th percentile of the sSFR for the galaxies with $10^{9.5} \mathrm{M_{\odot}} \leqslant M_{\ast} \leqslant 10^{10.5} \mathrm{M_{\odot}}$ at different redshifts.}
\label{fig:SFRevolution}
\end{figure}

We would like to address that ``quenched'', instead of being a galaxy type ``label'', is rather an evolutionary phase, which can be either a well established state on longer time scales or a transitional phase on shorter time scales. The former can correspond to present-day quenched elliptical galaxies, while the latter may often be seen in the history of present-day star-forming disks during their less active star-forming phases. This can be seen from Figure 3 of Paper I and Fig.~\ref{fig:SFRevolution} in this study, where the redshift evolution of a (present-day) star-forming disk galaxy and a quenched elliptical galaxy, respectively, are plotted in the $\log\,\mathrm{sSFR}-\log\,M_{\ast}$ plane, color-coded by redshifts. For the example present-day star-forming disk (Figure 3 in Paper I), it has experienced multiple star-forming episodes, interrupted by quenched phases, most noticeably at $z\sim 1$ and $z \sim 0.5$, where the $\rm sSFR$ drops significantly below a certain star-forming threshold and the central gas fraction $f_{\rm gas,\,<2R_{\rm hsm}}$ drops to below 5\%.  During these quenched phases, this system shows a deficit of cold and lower-angular momentum CGM gas. Luckily, each time a new episode of star formation manages to come back (once the feedback from the previous generation weakens) thanks to the resupply of lower-angular momentum CGM gas, as a perturbative consequence of galaxy interaction. In comparison, the present-day quenched early-type galaxy was not that ``lucky''. It was once an active star-forming galaxy before $z \sim 2$. It then became quenched and has remained quenched ever since, as indicated by a sudden drop of nearly three orders of magnitude in $\rm sSFR$ around $z\sim 1$.

\subsection{The role of angular momentum quenching}
\label{sec:AMQ}
The investigated effect on galaxy quenching in this study is essentially the same as what was termed as ``excess angular momentum inhibition'' or ``angular momentum quenching'', by a recent study of \citet{Peng_et_al.(2020)}. The proposed mechanism, strongly motivated by a bulge-quenching study from \citet{Renzini_et_al.(2018)} and recent HI observations (e.g., \citealt{Wang_et_al.(2016),Zhang_et_al.(2019),Murugeshan_et_al.(2019)}), was obtained through studying observed scaling relations. Within this framework, a galaxy first acquires and consumes lower-angular momentum gas to fuel its star formation. Once this is finished, due to either gas consumption by star-formation or low-angular-momentum gas removal by the AGN and stellar feedback (e.g. \citealt{Zjupa_et_al.(2017)}), high-angular-momentum CGM gas at the outskirts cannot further penetrate into the galaxy. As a consequence, the central star formation is then quenched.

It is worth noting that results from a recent study by \citet{Song21QuenchingInFilaments} using galaxies at $z=2$ from the HORIZON-AGN simulation (\citealt{Dubois14Horizon-AGNCosmicWeb}) are also in line with this study. The authors found suppressed star formation in galaxies living in a vorticity-rich environment by the edge of filaments, suggesting that less efficient gas transfer due to the high angular momenta inherited from their large-scale environment is responsible for quenching star formation.

We emphasize that ``angular momentum quenching'', instead of being a mechanism that quenches a galaxy, is rather a mechanism that keeps a galaxy quenched. This shall work together with other quenching mechanisms, for example, rapid gas consumption due to major mergers, AGN and/or morphology quenching etc, which produce ``a quenched galaxy'' in the first place. ``Angular momentum quenching'' may then kick in, preventing the CGM gas from flowing inwards to feed the central gas reservoir. As a result, further star formation is inhibited and the galaxy can remain quenched.

\subsection{The role of galaxy interaction}
In \citet{Peng_et_al.(2020)}, the kinematic effects that originate from the large-scale environment, such as galaxy mergers and fly-bys, were viewed as ``perturbations'' that trigger orbital instability and thus induce radial inflow of the (previously) high-angular momentum gas, which could then feed into the galaxy and sustain central star formation or rejuvenate a quenched galaxy (e.g., \citealt{Saintonge_et_al.(2011),Mancini_et_al.(2019)}). We would like to note that galaxy interactions under different angular momentum conditions may ultimately play different roles for star-forming disk galaxies and for quenched elliptical galaxies. In the former case, as is demonstrated in Paper I, the overall effects are to induce inflow of cold and lower-angular momentum gas. In the latter case, however, galaxy interactions in high-angular momentum environment shall be responsible for causing high angular momenta of the CGM gas and result in less efficient gas inflow, around present-day quenched galaxies.

\subsection{A mass dependence}
It is worth noting that below a stellar mass of $\sim 3\times 10^{10} \rm M_{\odot}$, the galaxy population is largely dominated by star-forming disks. In this case, where the CGM gas can effectively flow in, its angular momentum positively feeds back to the growth of the galaxy's stellar spin in the sense that the larger the former the higher the latter. This explains why \citet{DeFelippis_et_al.(2020)} found that the CGM specific angular momentum is higher in galaxies with higher stellar spins than with lower stellar spins within this mass range. This is fully consistent with our theoretical expectation regarding the connection between the CGM gas kinematics and galaxy stellar spins (e.g., see \citealt{Stewart(2017)} and \citealt{Tumlinson_et_al.(2017)}). Quenched early-type galaxies, however, are much more massive, for which the effects of the CGM spin modulation upon star formation are different from the previous case. This naturally renders the reported connections between the environmental angular momenta, the ambient CGM spins, galaxies' star formation rates, as well as the different evolution paths to a meaningful mass-driven result. As we go to higher stellar masses, the CGM gas angular momenta inherited from the larger-scale environment increase accordingly, and the ``angular momentum quenching'' mechanism becomes more significant. We further emphasise that such a mass-dependent effect can also originate in different initial conditions set at a much earlier stage.

\subsection{Origins of merging environments and fates of galaxies}
In this work, we have not studied the origin of the reported (potentially mass-dependent) environmental differences that are present across cosmic time, which have led to a diversity in galaxy morphology, star-forming activity and dynamical state among present-day galaxies. It is clear that the environmental differences must have been seeded in different large-scale torque fields that were fixed at a much earlier stage. This naturally suggests that the cosmological fates of present-day galaxies have already been determined by their large-scale environments at birth. More specifically, quenched early-type galaxies are not remnants of present-day star-forming disk galaxies, but rather have been set off on completely different evolutionary tracks due to different initial conditions at early epochs. This explicitly answers the ``nature versus nurture'' problem in the growths of galaxies. We encourage future studies investigating the link between the large-scale initial conditions and fates of present-day galaxies.

\section*{Acknowledgements}
It has been a fully enjoyable research experience with a fantastic collaboration working on this paper series. We would also like to thank Drs. Cheng Li, Caina Hao, Jing Wang, Lan Wang, Yougang Wang, Shihong Liao, Yong Shi, Xianzhong Zheng, Hong Guo, Yingjie Peng, and an anonymous referee for their very constructive and insightful suggestions and comments which improved the paper. This work is partly supported by the National Key Research and Development Program of China (No. 2018YFA0404501 to SM), by the National Science Foundation of China (Grant No. 11821303, 11761131004 and 11761141012). DX also thanks the Tsinghua University Initiative Scientific Research Program ID 2019Z07L02017. 

%%%%%%%%%%%%%%%%%%%%%%%%%%%%%%%%%%%%%%%%%%%%%%%%%%
\section*{Data availability}
General properties of the galaxies in the IllustrisTNG Simulation is available from \url{http://www.tng-project.org/data/}. The rest of the data underlying the article will be shared on reasonable request to the corresponding authors.

\bibliographystyle{mnras}
\bibliography{ref}

%\appendix
%\section{Circularity distribution of CGM}

%%%%%%%%%%%%%%%%% APPENDICES %%%%%%%%%%%%%%%%%%%%%

%%%%%%%%%%%%%%%%%%%%%%%%%%%%%%%%%%%%%%%%%%%%%%%%%%

\label{lastpage}
\end{document}